\documentclass[a4paper,fleqn,usenatbib]{mnras}

\usepackage{newtxtext,newtxmath}

\usepackage[T1]{fontenc}
\usepackage{ae,aecompl}

\usepackage{graphicx}	
\usepackage{amsmath}	
\usepackage{amssymb}	
\usepackage{booktabs}

\newcommand{\um}{$\mu$m}

\newcommand{\mstar}{M$_\ast$}
\newcommand{\lgmstar}{$\log($\mstar$/$M$_\odot)$}

\newcommand{\dindex}{D$_n$(4000)}
\newcommand{\nuvr}{NUV$-r$}
\newcommand{\hd}{H$\delta$}
\newcommand{\hda}{\hd$_A$}
\newcommand{\ewhda}{EW(\hda)}
\newcommand{\ha}{H$\alpha$}
\newcommand{\hae}{\ha}
\newcommand{\ewhae}{EW(\hae)}

\title[Molecular gas concentration, central SF, and bars]
{Linking bar- and interaction-driven molecular gas concentration with centrally-enhanced star formation in EDGE-CALIFA galaxies}

\author[R. Chown et al.]{Ryan Chown$^{1,2}$\thanks{Contact e-mail: \href{mailto:chownrj@mcmaster.ca}{chownrj@mcmaster.ca}},
Cheng Li$^{2}$\thanks{Contact e-mail: \href{mailto:cli2015@tsinghua.edu.cn}{cli2015@tsinghua.edu.cn}},
E. Athanassoula$^{3}$, Niu Li$^{2}$, Christine D. Wilson$^{1}$,
\newauthor
Lin Lin$^{4}$, Houjun Mo$^{2,5}$, Laura C. Parker$^{1}$, 
Ting Xiao$^{6}$
\\
$^{1}$Department of Physics and Astronomy, McMaster University, 1280 Main St. W., Hamilton, ON L8S 4L8, Canada\\
$^{2}$Tsinghua Center for Astrophysics and Physics Department, Tsinghua University, Beijing 100084, China\\
$^{3}$Aix Marseille University, CNRS, CNES, LAM, Marseille, France\\
$^{4}$Shanghai Astronomical  Observatory,  Nandan Road  80, Shanghai  200030, China\\
$^{5}$Department of Astronomy, University of Massachusetts Amherst, MA 01003, USA\\
$^{6}$Physics Department, Zhejiang University, Hangzhou 310058, China}

\date{Accepted XXX. Received YYY; in original form ZZZ}

\pubyear{2018}

\begin{document}
\label{firstpage}
\pagerange{\pageref{firstpage}--\pageref{lastpage}}
\maketitle

\begin{abstract}
We study the spatially resolved star formation history and  
molecular gas distribution of 58 nearby galaxies, using integral field spectroscopy
from the CALIFA survey and CO $J=1\rightarrow 0$ intensity mapping from the CARMA EDGE survey.
We use the 4000 \AA\ break (\dindex), the equivalent width of the H$\delta$ absorption line (\ewhda),
and the equivalent width of the H$\alpha$ emission line (\ewhae) to measure the recent star formation history (SFH) of these galaxies. 
We measure radial profiles of the three SFH indicators and molecular gas mass surface density,
from which we measure the level of centrally enhanced 
star formation and the molecular gas concentration.
When we separate our galaxies into categories of barred (17 galaxies), 
unbarred (24 galaxies), and merging/paired (17 galaxies) we 
find that the galaxies which have centrally-enhanced star formation (19/58)
are either barred (13/19) or in mergers/pairs (6/19) with relatively high molecular
gas concentrations.
A comparison between our barred galaxies and a
snapshot of a hydrodynamic $N$-body simulation of a barred galaxy 
shows that the current theory
of bar formation and evolution can qualitatively reproduce the main features
of the observed galaxies in our sample, including both the sharp decrease of 
stellar age in the galactic center and the gradual decrease of age with
increasing distance from center. These findings provide substantial evidence for 
a picture in which cold gas is transported inward 
 by a bar or tidal interaction,
which leads to the growth and rejuvenation of star formation in the central region.
\end{abstract}




\section{Introduction}
\label{sec:introduction}

Bars play an essential role in the secular evolution of galaxies.
Simulations have shown that the growth of a bar 
causes 
gas to either form a ring structure or 
fall inwards and trigger central star formation \citep{athanassoula1992b, athanassoula1994, piner1995,  athanassoula2013, sormani2015}.
Minor mergers and tidal interactions have a similar effect as bars, 
as these events also tend to drive molecular gas 
inward \citep{barnes1991}.
The subsequent star formation from these processes leads to the growth of 
the central disky pseudobulge \citep[][and references therein]{kormendy2004, athanassoula2005}.
Internal processes can quench star formation, such as feedback 
from an active galactic nucleus (AGN), or the growth of the central bulge 
which can stabilize the gas disk \citep{martig2009}.
Bars can counteract these quenching mechanisms by transporting gas to the center
which can fuel subsequent central star formation.

The most commonly-used tracer of molecular gas mass in the interstellar medium is line emission of the CO molecule, e.g. CO $J=1\rightarrow 0$, $J=2\rightarrow 1$, etc. \citep[][and references therein]{bolatto2013}.
Observational studies of CO have found elevated molecular gas
concentrations in barred galaxies compared to their unbarred counterparts \citep[e.g., ][]{sakamoto1999, sakamoto2000, jogee2005, sheth2005, regan2006, kuno2007}.
It is also known that star formation rates (SFRs)
are higher in the central region of barred galaxies compared to unbarred galaxies
\citep[e.g, ][]{hawarden1986, devereux1987, puxley1988, ho1997, ellison2011, oh2012, zhou2015}.
Minor mergers and galaxy-galaxy interactions have also been found to correlate with increased 
central star formation \citep[e.g., ][]{li2008, ellison2011, wang2012, lin2014}. 
Interaction-induced enhancement of star formation is found mainly in galaxies with projected separations less than 
$\sim100$ kpc \citep{li2008, patton2013, ellison2013}.
A number of recent studies have used both molecular gas tracers and star formation
indicators to study central star formation and cold gas in interacting galaxies \citep[e.g., ][]{saintonge2012, stark2013, kaneko2013, violino2018}, 
finding lower gas depletion times and enhanced gas content in these galaxies. Galaxies in dense environments also tend to have more centrally-concentrated molecular gas and enhanced star formation \citep[e.g., ][]{mok2017}.

Most earlier optical studies of galaxies used single-fiber measurements from 
the Sloan Digital Sky Survey \citep[SDSS; ][]{york2000}. Integral-field unit (IFU) surveys such as the Calar Alto Legacy Integral Field Area (CALIFA) 
survey \citep{sanchez2012, walcher2014, sanchez2016}, the
SDSS-IV Mapping nearby Galaxies at Apache Point Observatory Survey \citep[MaNGA; ][]{bundy2015, blanton2017}, and
the Sydney-AAO Multi-object Integral field spectrograph \citep[SAMI; ][]{croom2012} have provided spatially resolved spectroscopy for thousands of galaxies in the local Universe, enabling detailed studies of the correlation of internal structure of galaxies with their star formation properties. 

Of particular interest for the present work, \citet{lin2017a} analyzed 57 nearly face-on spiral galaxies using CALIFA IFU data.
They measured the recent star formation history (SFH) using three
parameters extracted from the CALIFA data: the 4000~\AA-break \dindex, and the equivalent widths (EW) of the H$\alpha$ emission line $\log$\ewhae~and 
H$\delta$ absorption line \ewhda.
A considerable fraction of their galaxies (17/57) had a central drop (``turnover'') in the \dindex, and a central upturn
in $\log$\ewhae~and \ewhda, indicating recent star formation in the center.
Interestingly, almost all of these ``turnover'' galaxies are barred, while only half of the barred galaxies in their sample present a turnover feature, suggesting that a bar is a \textit{necessary but 
not sufficient} condition for central star formation enhancement.
The only parameter found to be correlated with the level of central star formation is the bar length, an indicator of bar strength.
Together with the results of \citet{kuno2007}, for example, who found 
a correlation between bar strength and molecular gas concentration,
one might expect enhanced central star formation to be associated with 
molecular gas concentration. Observations of the cold gas within galaxies, with spatial resolution comparable to the optical IFU data, are needed in order to clearly examine the correlation of the two components.

Uniform samples of high-sensitivity (detections of $\Sigma_\mathrm{H_2}\sim$ 1 $M_\odot$ pc$^{-2}$), high spatial resolution (sub-kpc) cold gas measurements of nearby galaxies are available, however 
the sample sizes range from a few up to 30-50 galaxies.
Although a sample of 30-50 galaxies is sufficient for many purposes, the effective sample size can quickly become much lower after selection cuts (on redshift, stellar mass, etc.) and/or dividing the galaxies into different categories for comparison (e.g. barred or unbarred).
Some notable studies and surveys are \citet{kennicutt2007}, \citet{bigiel2008}, 
the \ion{H}{i} Nearby Galaxy Survey \citep[THINGS; ][]{walter2008}, 
and the HERA CO-Line Extragalactic Survey \citep[HERACLES; ][]{leroy2009}. At slightly lower spatial resolution, recent surveys have obtained spatially resolved CO spectra for significantly larger samples, such as the James Clerk Maxwell Telescope (JCMT) Nearby Galaxies Legacy Survey \citep[NGLS; 155 galaxies, $\sim$50\% detected; ][]{wilson2012}, the Combined Array 
for Research in Millimeter-wave Astronomy (CARMA) Extragalactic Database for Galaxy Evolution (EDGE) 
CO survey \citep[126 galaxies, 82\% detected; ][]{bolatto2017},
and the CO Multi-line Imaging of Nearby Galaxies survey (COMING; 127 galaxies; Sorai, K., et al. 2018, in preparation). Galaxies in the CARMA EDGE survey were selected from the CALIFA survey, and 
were observed in CO $J=1\rightarrow 0$ with a similar field-of-view and angular resolution ($\sim4.5$\arcsec) as the CALIFA data ($\sim 2.5$\arcsec).
The similar resolution and field-of-view were intended to enable joint analyses of the stellar populations and cold gas content of nearby galaxies.
Recent work by \citet{utomo2017} found, using EDGE and CALIFA data, that barred and interacting galaxies tend to have smaller center-to-disk gas depletion time ratios than unbarred, isolated galaxies.

In this paper we extend the work of \citet{lin2017a} by explicitly linking the central SFH and resolved 
gas properties in barred, unbarred, and interacting galaxies. 
We have used spatially resolved maps of CO $J=1\rightarrow 0$ from EDGE and 
SFH indicators from CALIFA to accomplish this goal. 
We find that molecular gas concentrations are indeed associated with
centrally-enhanced star formation, and this link is seen in both barred galaxies 
and interacting galaxies. Our main result is that the level of centrally-enhanced star formation in 
barred galaxies is positively correlated with 
molecular gas concentration (correlation coefficient $r=0.64$), while unbarred galaxies show little-to-no centrally-enhanced star formation and no correlation ($r=0.09$). Some merger/pair galaxies have centrally enhanced star formation, but the correlation between the level of enhanced star formation and gas concentration is weak ($r=0.26$).
In addition, we have compared these observational results with an $N$-body simulation 
of the gas and stellar distributions of a barred galaxy. The similarities between 
the simulation and the real galaxy suggest that the current theory of 
bar formation can qualitatively reproduce the key features of real galaxies.

The structure of this paper is as follows. 
In \S\ref{sec:data} we describe the data used and how they are processed.
We present our observational results in \S\ref{sec:obsresults} and the 
comparison with the $N$-body simulation in \S\ref{sec:simresults}.
We discuss some questions in light of our results and highlight interesting individual galaxies in \S\ref{sec:discussion}.
Finally, we summarize our work in \S\ref{sec:conclusions}. Tables of galaxy properties and a discussion of a few unusual galaxies are given in the Appendix. Throughout this paper we 
assume a $\Lambda$CDM cosmology with parameters $H_0=67.7$ km s$^{-1}$ Mpc$^{-1}$, 
$\Omega_\mathrm{\Lambda}=0.693$, and $\Omega_\mathrm{m,0}=0.307$, following the 
results from the \textit{Planck} satellite \citep{planck15-7}. 

\begin{figure*}
\centering
\includegraphics[width=0.4\textwidth]{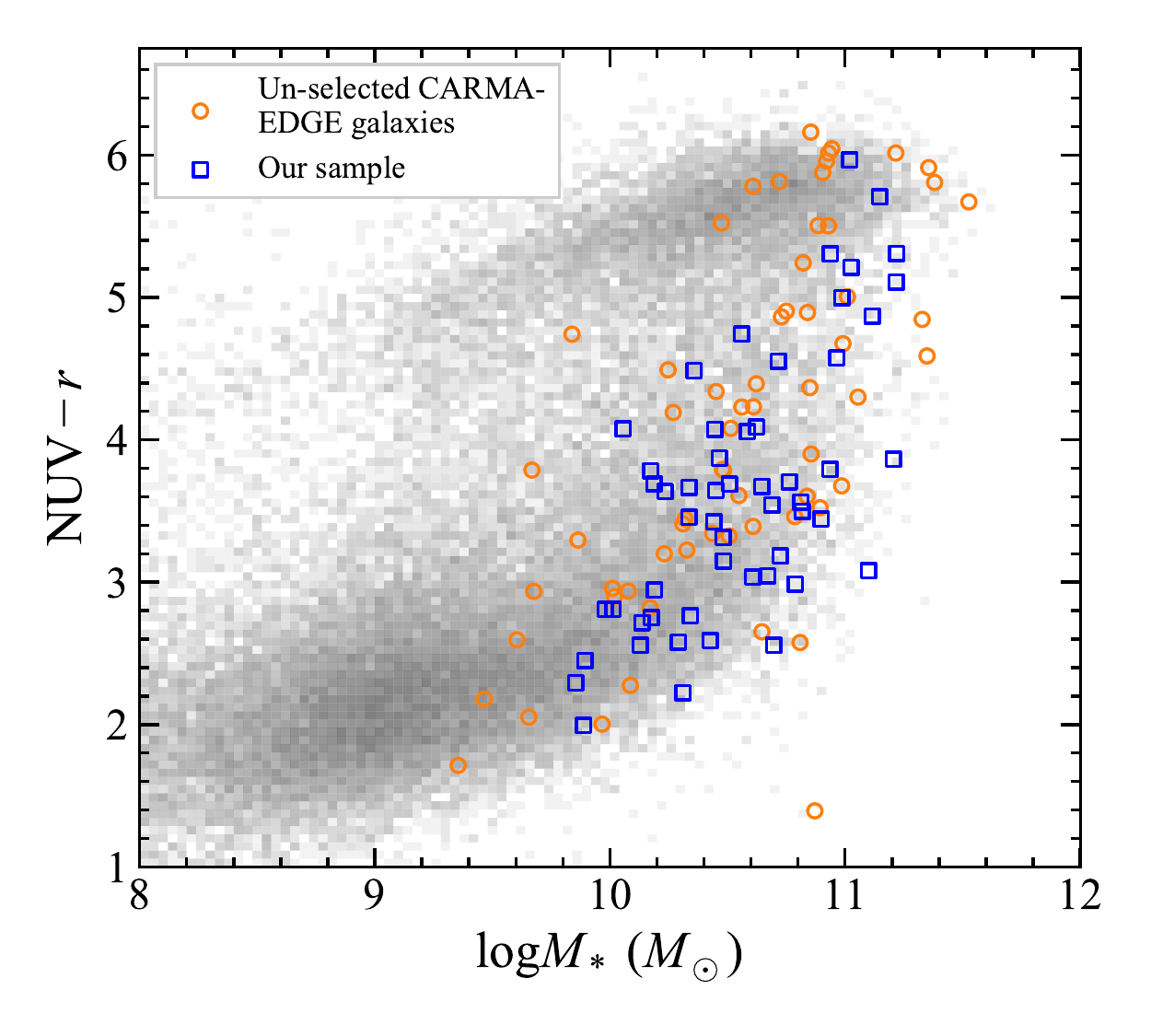}\includegraphics[width=0.4\textwidth]{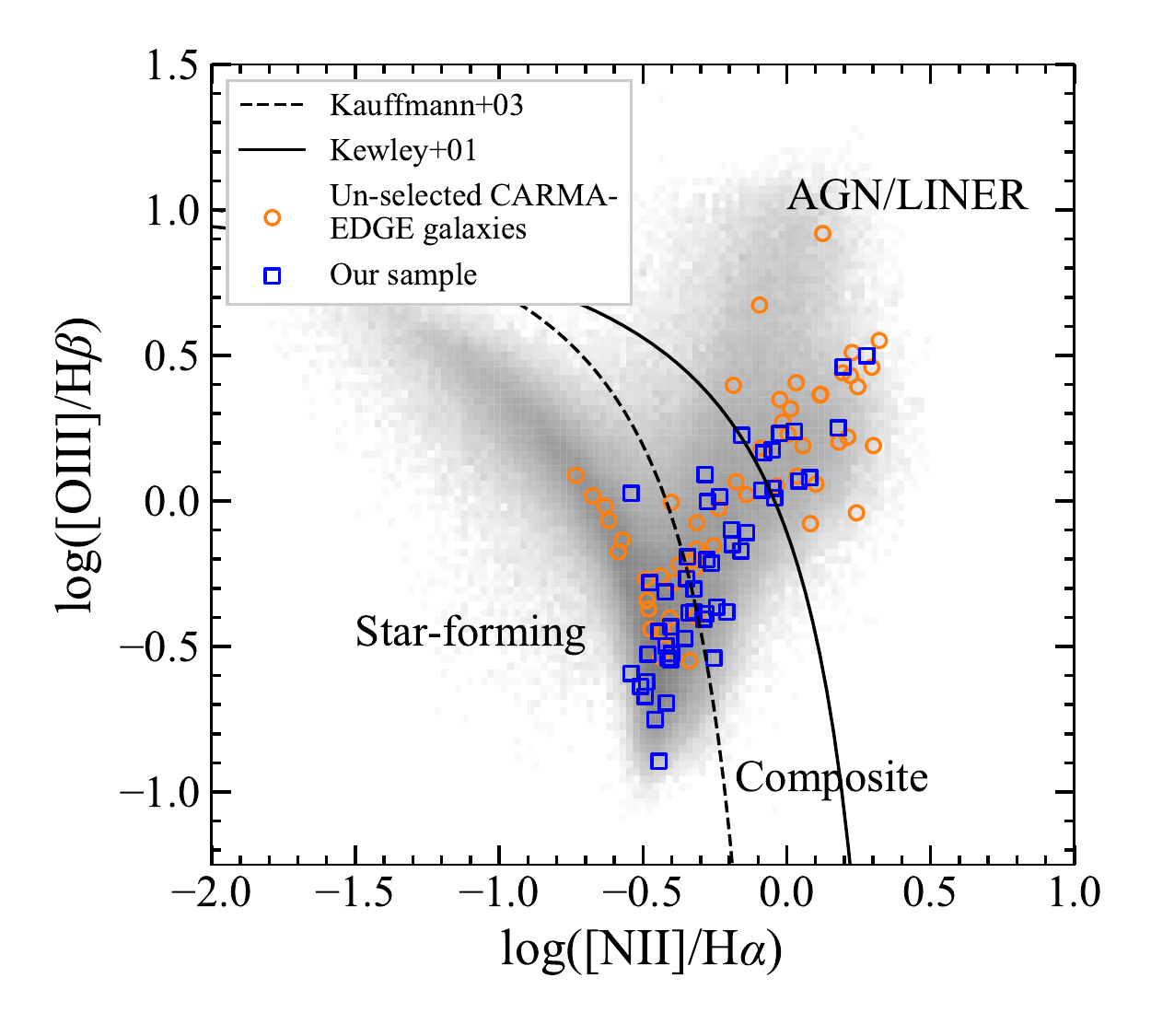}
\caption{\textit{Left:} \nuvr~vs. \lgmstar~for the sample  used in this work (blue squares), the remaining galaxies in the CARMA EDGE survey (orange circles), and a volume-limited sample of low-redshift galaxies with $0.003\leq z\leq 0.03$ from the NASA-Sloan Atlas (grey). \textit{Right:} our sample and the unselected CARMA EDGE galaxies 
shown on the BPT diagram, overlaid on a volume-limited sample selected from the MPA-JHU catalog (grey). 
Points lying between the lines of \citet{kewley2001} and \citet{kauffmann2003} are composites, while points toward the lower left are star-forming, and points toward the upper right are LINER.
These figures show that our sample consists of mostly 
star-forming galaxies with stellar masses above $\sim10^{10} M_\odot$, and are mainly star-forming/composite according to the BPT diagram. 
We use the spatially-resolved BPT diagrams for each of our galaxies to exclude spaxels from our analysis which are classified as composite or LINER. 
Note that some galaxies are excluded from the left panel if they do not have NUV data in the NSA, and some galaxies are excluded from the right panel if the signal-to-noise is less than 3.0 in any of the relevant emission lines.}
\label{fig:sample1}
\end{figure*}

\section{Data and processing}
\label{sec:data}

\subsection{The CARMA EDGE and CALIFA surveys}

CARMA EDGE \citep{bolatto2017} is a survey of CO emission in 126 nearby galaxies carried out using the 
CARMA interferometer \citep{bock2006}. 
The CARMA EDGE sample was selected from the CALIFA sample with high 
fluxes in the 22\um\ band from the 
\textit{Wide-field Spectroscopic Explorer} (\textit{WISE}) survey.
The requirement for high mid-infrared flux means that the sample is 
mainly gas-rich and actively star-forming, 
given the correlation between the mid-infrared luminosities from 
\textit{WISE} and the molecular gas mass \citep[e.g.][]{jiang2015}.
The sample consists of galaxies 
imaged in $^{12}$CO and $^{13}$CO with sensitivity, angular resolution and 
field-of-view well-matched to CALIFA data.
The typical $3\sigma$ molecular gas mass surface density sensitivity is 11 $M_\odot \mathrm{pc}^{-2}$, and the typical angular resolution is 4.5\arcsec\ \citep{bolatto2017}.
We use the publicly available 
$^{12}$CO $J=1\rightarrow 0$ integrated flux maps from CARMA EDGE. 
Specifically, we chose the version of these maps made
by creating a mask using a smoothed version of the data cubes, 
and applying this mask to the original (un-smoothed) cubes before integrating.
The maps are sampled with 1\arcsec\ $\times$ 1\arcsec\ pixels.

We use optical IFU data from the 3rd data release (DR3) of the CALIFA survey \citep{sanchez2012}.
The CALIFA survey consists of about 600 galaxies observed with the 
PMAS/PPak integral-field spectrograph at the Calar Alto Observatory \citep{roth2005, kelz2006}.
 The CALIFA datacubes are available
in three spectral setups: 
(1) a low-resolution setup with 6 \AA~spectral resolution ($V_{500}$),
(2) a medium-resolution setup with 2.3 \AA~resolution ($V_{1200}$),
and (3) the combination of $V_{500}$ and $V_{1200}$ cubes (called COMBO).
The mean angular resolution of CALIFA is 2.5\arcsec, which is similar to SAMI and MaNGA.
The maps are sampled with 1\arcsec\ $\times$ 1\arcsec\ pixels.
We used the COMBO data cubes from CALIFA DR3 where available. For the 8 CARMA EDGE galaxies with no COMBO datacubes available, we used the $V_{500}$ datacubes instead.

\begin{figure*}
\centering
\includegraphics[width=0.9\textwidth]{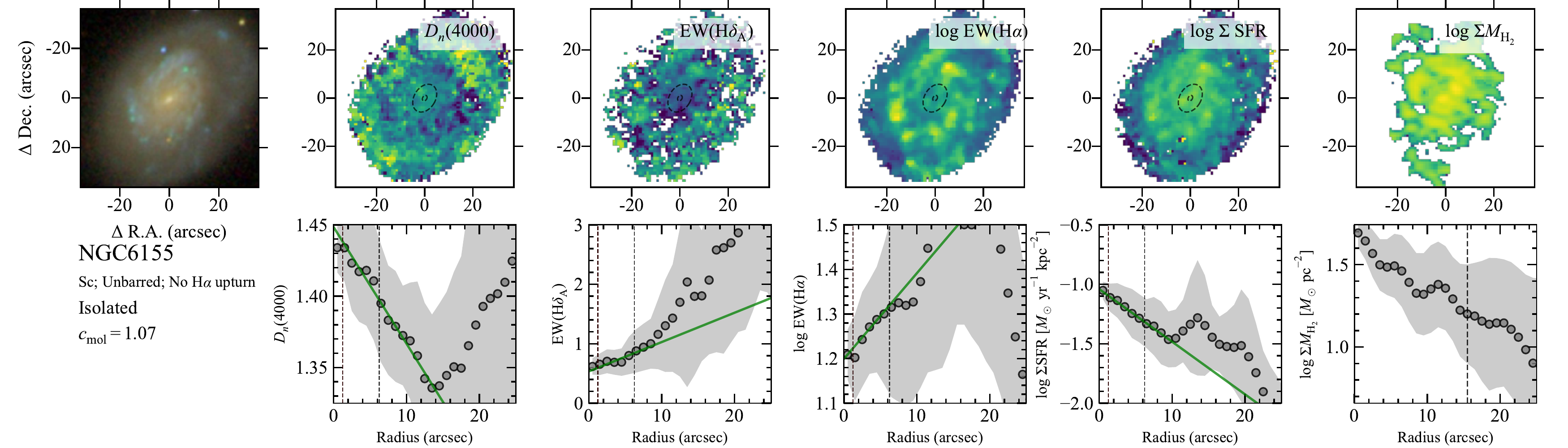}
\includegraphics[width=0.9\textwidth]{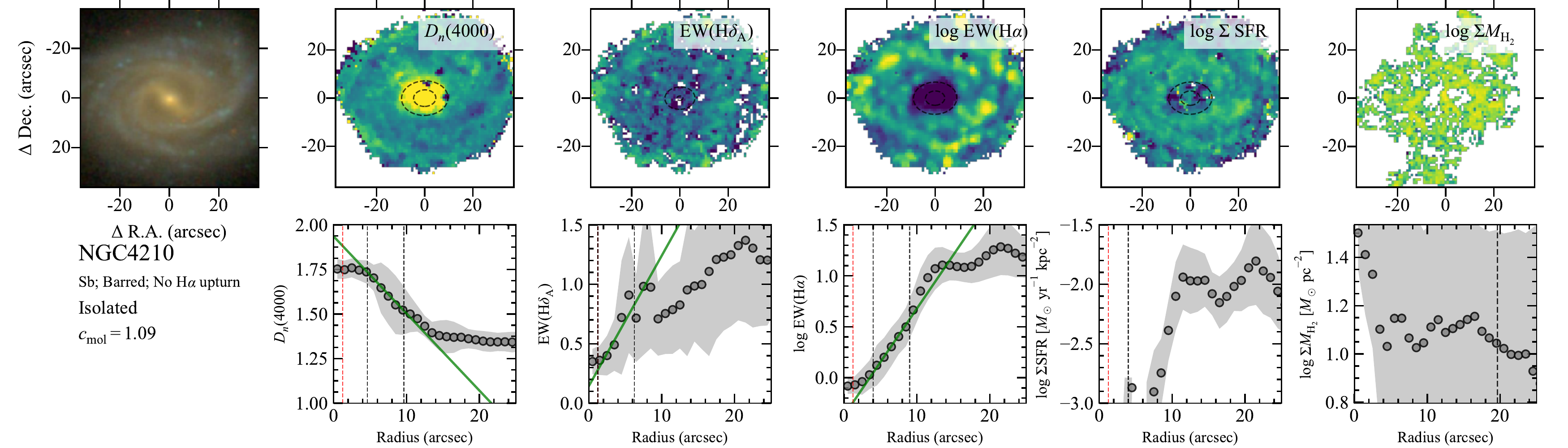}
\includegraphics[width=0.9\textwidth]{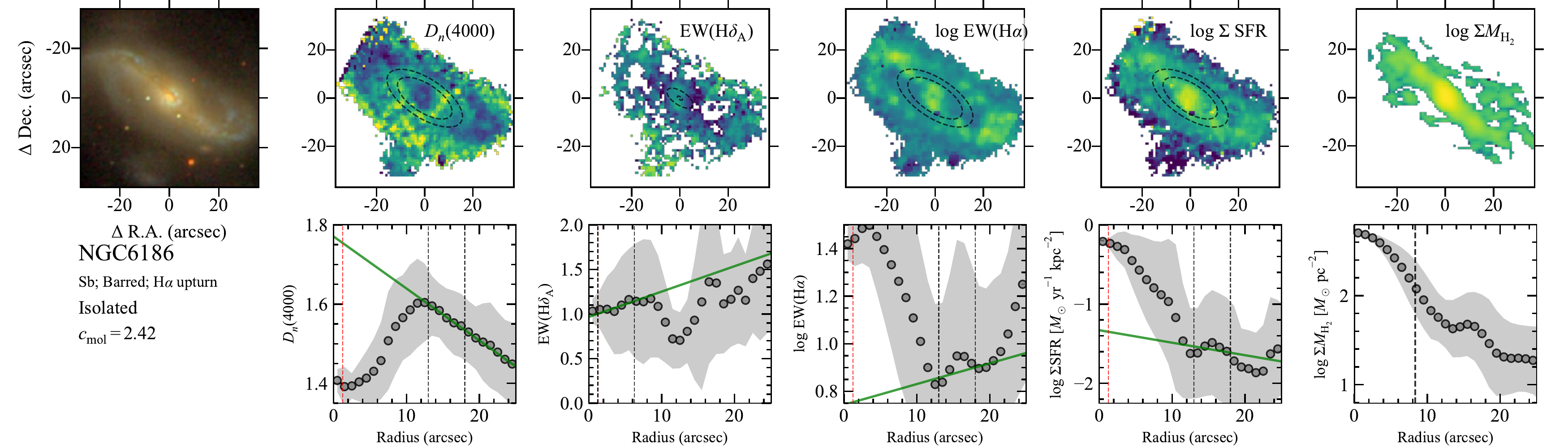}
\includegraphics[width=0.9\textwidth]{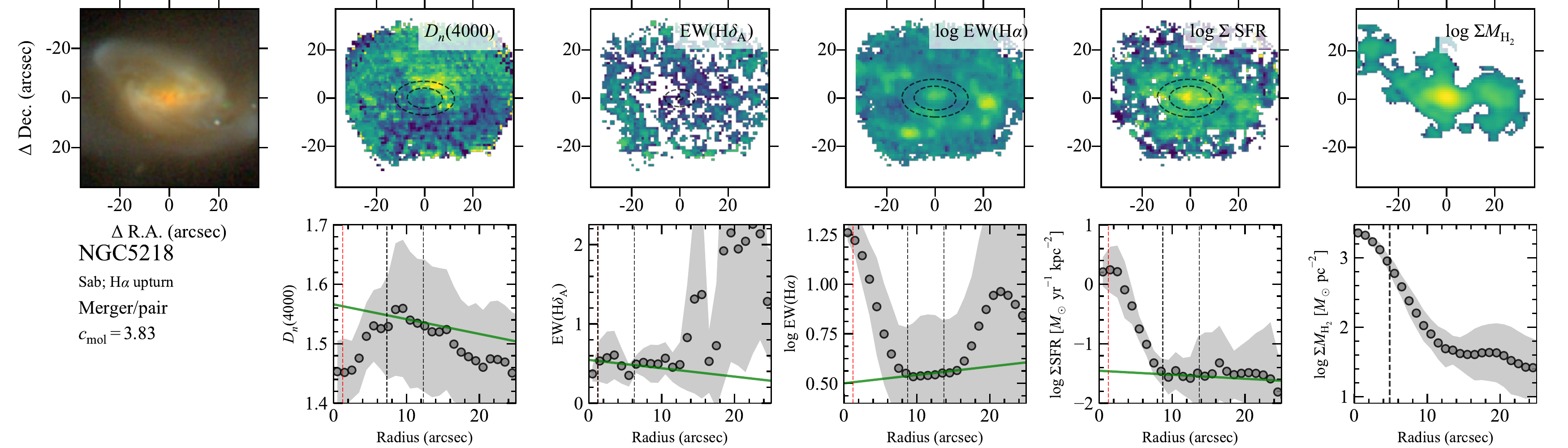}
\caption{Maps and radial profiles for example galaxies in our sample. 
For each galaxy, the upper and lower panels on the left show 
the optical \textit{gri} image from SDSS,
the Hubble type and subtype (from CALIFA), 
an indication of whether it has a central upturn in $\log$\ewhae~ 
(described in Sec.~\ref{sec:turnoverid}),
and isolated or merger/pair status (described in Sec.~\ref{sec:morphclass}). 
The maps and radial profiles of \dindex, \ewhda, log\ewhae,  $\log\Sigma_\mathrm{SFR}$, and $\log\Sigma_\mathrm{H_2}$ are shown on the right.
On each image or map, ellipses are plotted with semi-major axis of the inner and outer  
regions where the linear fits are performed.
The vertical lines in the radial profiles match the corresponding ellipses. 
The vertical line in the $\Sigma_\mathrm{H_2}$ profile is the half-gas-mass radius, which is used to calculate gas concentration; the value of the concentration index is indicated below the \textit{gri} image.
The innermost vertical black lines in the middle four panels are determined by eye for each quantity
separately (\dindex, \ewhda, $\log$\ewhae), and the outermost lines are placed 
5\arcsec\ beyond these lines.
The fitting regions are the same for $\log$\ewhae~and $\Sigma_\mathrm{SFR}$. 
Any peaks/troughs in the radial profiles at larger radii than where the fitting was performed 
(e.g. at $\sim 12$ \arcsec\ in \dindex\ for NGC 6155)
were identified with disk-like structure in the optical images, and so are not classified as the 
turnover/upturn radii.
In the 4 middle radial profiles, if the CALIFA beam half-width 
at half maximum is smaller than the inner fit radius, it is shown as the red dashed line.
The green lines show the linear fits which are done between 
the two vertical black lines. 
If no green lines are shown in $\Sigma_\mathrm{SFR}$, the signal-to-noise was insufficient to fit a line to the radial profile.
More details can be found in the text.}
\label{fig:mapsandprofiles1}
\end{figure*}

\subsection{Sample selection}
\label{sec:sample}

Starting with all 126 CARMA EDGE galaxies, we remove galaxies which are close to edge-on
by requiring the minor-to-major axis ratio to be $b/a>0.3$.
For our analysis we need to be able to measure radial profiles of 
the molecular gas mass density, so we exclude galaxies which are not detected 
in CO emission, which would prevent us from measuring a radial profile. 
We use same definition of ``detection'' as \citet{bolatto2017}, namely 
a cube which has at least one beam with at least a 5$\sigma$ detection in at least
one velocity bin ($\delta v = 10$ km/s). 
Furthermore, we remove four galaxies from our sample which are classified as Seyferts according to the SIMBAD database,
since we suspect such galaxies may have different central star formation properties (and perhaps molecular gas properties) than non-AGN. 
We remove an additional 6 galaxies which we classify as AGN based on their BPT diagrams.
We would like to emphasize that the correlation of nuclear activity with cold gas concentration is an interesting topic by itself, but this is outside the scope of the current paper.

These requirements leave us with 58 galaxies. 
Some basic properties
of the sampled galaxies are listed in Table~\ref{tab:tableA1}.
The galaxies in our sample are located between 24 and 128 Mpc away, or 68 Mpc on average.
The angular resolutions of CALIFA ($\sim$2.5\arcsec) and CARMA ($\sim$4.5\arcsec) at these 
distances correspond to physical diameters of 0.27-1.5 kpc and 0.5-2.8 kpc respectively.
The pixel scale of 1\arcsec\ corresponds to a physical scale of 0.1-0.6 kpc.

Figure~\ref{fig:sample1} shows our sample (blue squares) in the \nuvr\ vs. \lgmstar~plane
on top of a volume-limited sample ($0.003\leq z \leq 0.03$) selected from the NASA-Sloan Atlas\footnote{NSA: \url{http://www.nsatlas.org}} (NSA), version 0.1.2. The NUV magnitudes in the NSA catalog are 
from the \textit{Galaxy Evolution Explorer} (\textit{GALEX}) \citep{martin2005}, and the stellar masses are estimated by \citet{blanton2007} based on SDSS \textit{ugriz} Petrosian magnitudes; 
see \citet{blanton2005a, blanton2005, blanton2011} for details on the NSA. As can be seen, 
our sample covers a wide range of global properties from the star forming main sequence (the lower part of the plane) through the green valley and into the red sequence,
and is roughly representative of the CARMA EDGE survey (the orange circles plus our sample).
Compared to the volume-limited sample, galaxies in the EDGE sample 
have relatively high stellar masses with $M_\ast\ga 3\times10^{9} M_\odot$, and are predominately blue or green in colour with some extending into the red sequence.
Their mostly-blue \nuvr~colours (indirectly) suggests that they 
have mainly high molecular gas mass fractions 
according to the tight empirical relation between H$_2$ mass fraction and \nuvr~\citep{saintonge2017}. This is expected, given the requirement for high 22-$\mu$m \textit{WISE} luminosity 
in the EDGE sample selection.

Figure~\ref{fig:sample1} also shows our sample on the Baldwin, Phillips and Terlevich (BPT) diagram \citep{baldwin1981}. For reference, we show 
the BPT diagram for a volume-limited sample from the MPA-JHU catalog 
derived from SDSS DR8 data. For this plot we have measured 
the fluxes of the four emission lines in the central 3\arcsec\ of all CARMA 
EDGE galaxies using the processed CALIFA data. 
Our galaxies fall mainly in the star-forming and composite regions, with some extending to the LINER region. 
Compared to the full EDGE sample (orange + blue), our sample lacks
Seyfert galaxies which is a consequence of our sample selection as 
described above. Note that some galaxies are not shown on this diagram, 
because the signal-to-noise is required to be greater than 3.0 in all four emission lines.

In a later section we analyze the gas content and star formation rate surface density in the nuclei of our galaxies. For that analysis we require at least 50 percent of the pixels in the central 500 pc (radius) to be detected in CO, and not be classified as LINER. 
31/58 galaxies from our primary sample satisfy these more stringent requirements. This smaller sample (which we call our ``reduced'' sample) is used in \S\ref{sec:profiles}.

\subsection{Morphological classification}\label{sec:morphclass}

We visually classify all of the galaxies in our sample  (done by the first two authors) as either barred or unbarred, 
and we cross-check our results with two or three independent classifications from the literature. First, we use the morphological classification by the CALIFA team \citep{walcher2014}, 
who classify the galaxies as barred, unbarred, or uncertain.
The CALIFA morphological classification is done by-eye by the CALIFA collaboration using \textit{r}- 
and \textit{i}-band SDSS DR7 images, as described in \citet{walcher2014}.
Next, we cross-match our sample with the SIMBAD database \citep{wenger2000} to get morphological types and references for each. 
Additionally, there are 19 galaxies in our sample that overlap with the sample of \citet{lin2017a}, who performed a reliable bar classification by
applying the IRAF task {\tt ELLIPSE} to the background-subtracted $r$-band images from SDSS. 
For most galaxies in our sample, these two (or three where available) cross-checks on the bar status agree. Our final classification of barred or unbarred
is our best judgment of the CALIFA, SIMBAD, \citet{lin2017a} and our own by-eye classification.

\begin{figure*}
\centering
\includegraphics[width=0.9\textwidth]{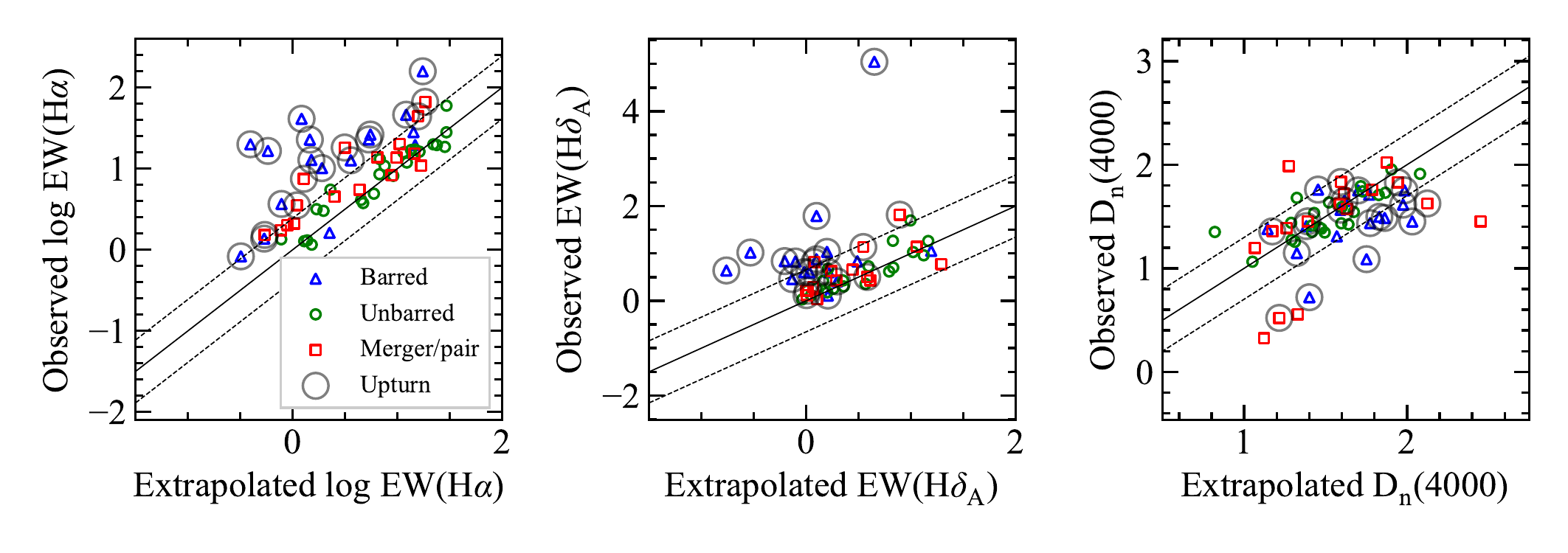} 
\caption{Comparison between the observed central value and extrapolated central value of the three star formation history indicators used. In each panel, a 1:1 relation and the 1$\sigma$ scatter around this relation (not the scatter about the mean) are shown. As in \citet{lin2017a}, we define ``upturn'' galaxies as those 
which lie above the $+1\sigma$ line in the left panel. Upturn galaxies are indicated with grey circles.}
\label{fig:obs_extrap_scatter}
\end{figure*}

Next, each galaxy is classified as either isolated or interacting 
with another galaxy (mergers or pairs). 
Galaxies which are classified by the CALIFA team 
as mergers or isolated are initially put into these two categories.
There are a small number of galaxies with uncertain merger status based on their classification.
We then cross-check all isolated/merger classifications 
with the SIMBAD database, which resolves the uncertain cases, 
and moves some galaxies classified as isolated into the paired category. 
Galaxies in SIMBAD are classified as interacting if they belong to
catalogues of interacting galaxies such as \citet{vorontsov2001};
 galaxies are classified as pairs if they belong to any of the 
available catalogues of paired galaxies 
\citep[e.g., ][]{karachentsev1972, turner1976, barton2003}. 

We visually examine the SDSS \textit{gri} images of all galaxies classified as pairs, and in a small number of cases, the companion galaxies are too 
far away to affect the central star formation (greater than $\sim 200$ kpc as discussed in \S\ref{sec:introduction}, and/or at significantly different redshifts). Such cases are moved into either the isolated barred or isolated unbarred category.

These classifications are used to group the galaxies into 
three categories: isolated barred, isolated unbarred, and merger/pair/interacting. Note that pair galaxies may be barred or unbarred.
In summary, we have 17 isolated barred galaxies, 24 isolated unbarred galaxies, 
and 17 merger or interacting pair galaxies.
The reduced sample mentioned in the previous section consists of 11 barred galaxies, 13 unbarred galaxies, and 7 merger/pair galaxies.

The means and standard deviations of the distances of the barred, unbarred and merger/pair categories are
$73 \pm 32$ Mpc, $66 \pm 27$ Mpc, and $66 \pm 24$ Mpc, respectively. Given
the similarity of these distributions, we do not expect any distance-related
biases to affect the physical resolution of our data.
Furthermore, we compare the populations using distance-independent quantities.

\begin{figure*}
\centering
\includegraphics[width=0.8\textwidth]{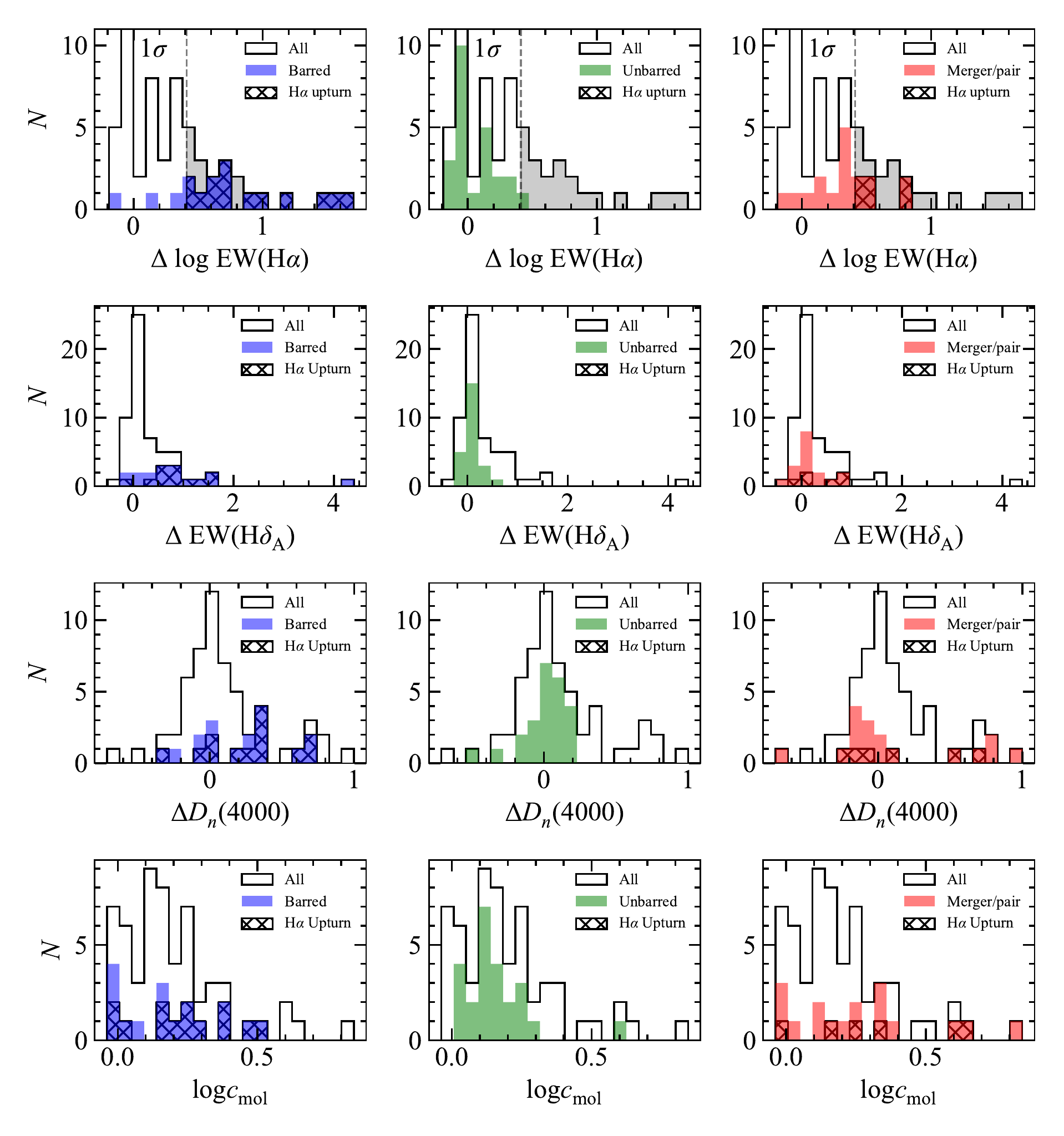} 
\caption{Panels from top to bottom are histograms of the $\log$\ewhae~upturn strength, \ewhda~upturn strength, \dindex~turnover strength, and $\log c_\mathrm{mol}$ for barred (blue), unbarred (green) and merger/pair galaxies (red). In each row, the histogram of the corresponding parameter for the full sample is plotted and repeated 
in the three panels as the black histogram. 
In the top row the vertical line indicates the 1$\sigma$ scatter (0.410 dex)
of all the points from the 1:1 relation of the observed value of log \ewhae\ at the center to the extrapolated value. Galaxies 
with $\Delta\log$\ewhae\ $\geq\sigma$ (those lying above the $+1\sigma$ line in Fig.~\ref{fig:obs_extrap_scatter}) are classified as having a central 
upturn (shown in the grey histogram in the top panels). 
The distribution of the upturn galaxies in each panel is plotted as the hatched histogram. This figure immediately shows that the upturn galaxies
are either barred or mergers/pairs, and none of the unbarred galaxies 
have an $\log$\ewhae~upturn. 
}
\label{fig:tstrength_hists}
\end{figure*}

\subsection{Maps and radial profiles of recent SFH diagnostics} \label{sec:spectralfitting}

We measure three spectral indices, which together tell us
the recent SFH for a given spaxel of our galaxies \citep{bruzual2003, kauffmann2003a, li2015, lin2017a}:
\begin{description}
    \item[\ewhae---] Equivalent width (EW) of the H$\alpha$ emission line. This quantifies the ratio of the current star formation rate \citep[0-30 Myr; ][]{kennicutt2012} to the recent past star formation rate.
    \item[\ewhda---] EW of the H$\delta$ absorption line (the subscript indicates absorption). 
    A strong H$\delta_\mathrm{A}$ line indicates a burst of star formation which ended 0.1 to 1 Gyr ago \citep{kauffmann2003b}.
    \item[\dindex---] The 4000 \AA~break. This index is sensitive to stellar populations formed 1-2 Gyr ago. In practice, if \dindex~$\lesssim 1.6$, there has been star formation in this time frame \citep{li2015}.
\end{description}

For each galaxy in our sample, we perform full spectral fitting to each spaxel in the CALIFA DR3 COMBO data cube
(or $V_{500}$ for the few galaxies which do not have COMBO datacubes), using the Penalized Pixel-Fitting code\footnote{\url{http://www-astro.physics.ox.ac.uk/~mxc/software}} (pPXF) \citep{cappellari2004, cappellari2017}.
The \citet{bruzual2003} simple stellar populations (SSP)
and the \citet{calzetti2000} 
reddening curve were used during the fitting.
The result of the full spectral fitting is a best-fit model spectrum representing the stellar component of the spaxel (continuum plus absorption lines) which is a linear combination of the SSPs, 
plus a color excess $E(B-V)$ quantifying the overall dust extinction. Both the observed spectrum and the model spectrum are corrected for dust extinction according to this $E(B-V)$. The equivalent width of 
H$\delta_\mathrm{A}$ and \dindex~are then measured from the model spectrum. In addition, we obtain the luminosity-weighted
age of the spaxel based on the luminosity and coefficient of the SSPs that form the best-fit spectrum.

The model spectrum is then subtracted from the observed one, and we measure both the dust-corrected flux and the equivalent width for the H$\alpha$ emission line. We correct for the dust extinction 
based on the Balmer Decrement measured from the observed spectrum. For this we have assumed a temperature of $10^4$ K, an electron density of $10^2$ cm$^{-3}$ in the \ion{H}{ii} regions, an intrinsic H$\alpha$-to-H$\beta$ flux ratio of 2.86 in case-B recombination \citep{osterbrock2006}, as well as a \citet{calzetti2000} reddening curve. The dust-corrected H$\alpha$ flux is converted to a luminosity using the distance assuming the adopted cosmological parameters  \S\ref{sec:introduction}. A star formation rate (SFR) is then
estimated by multiplying the H$\alpha$ luminosity by 
$5.3\times10^{-42}M_\odot$ yr$^{-1}$ (erg s$^{-1}$)$^{-1}$ as 
in \citet{murphy2011, hao2011, kennicutt2012}. This SFR 
calibration adopts the stellar initial mass function (IMF) 
from \citet{kroupa2003}.

We compute radial profiles of each quantity by azimuthally averaging  
the maps in elliptical annuli separated by 1\arcsec\ along the 
semi-major axis.
The position angles and minor-to-major axis ratios are taken from the 
CALIFA DR3 supplementary tables
\citep{walcher2014}.
Partial pixel overlap within each annulus 
is taken into account, and pixels with signal-to-noise ratio less than
1 are set to zero in the averages.
Using a $1\sigma$ value instead of zero does not significantly affect the radial profiles at the radii used in our analysis.
The SDSS {\it gri} images and the processed maps of \dindex, \ewhda\
and $\log$\ewhae, as well as their radial profiles are shown in 
Figure~\ref{fig:mapsandprofiles1} for four example galaxies 
from our sample: an unbarred 
galaxy in the top (NGC6155), followed by two
barred galaxies (NGC4210 and NGC6186) and a merging galaxy (NGC5218).
The first two galaxies present similar radial profiles in the recent 
SFH indicators in the sense that, from galactic center to larger radii,
\dindex\ decreases while both $\log$\ewhae\ and \ewhda\ increase. 
This radial profile shape indicates a relatively old stellar population
in the inner region, and less star formation in the recent past in this
region. Large samples of IFU observations such as MaNGA have shown that
radial profiles like this are typical for the general population of
galaxies in the local Universe, although the amplitudes and slopes of
the profiles depend on galaxy stellar mass \citep[e.g.][]{wang2018}. 

Different from the top two galaxies in Figure~\ref{fig:mapsandprofiles1},
the bottom two galaxies in the same figure show a significant upturn 
in $\log$\ewhae\ and/or a significant turnover in \dindex\ in their innermost
region, indicating star formation has been recently enhanced in the central region. 
Galaxies with a central turnover in \dindex\ were called ``turnover'' galaxies in \citet{lin2017a}. Those authors found that almost all  
turnover galaxies are barred. We note that the two galaxies in Figure~\ref{fig:mapsandprofiles1} with 
a turnover feature are a barred galaxy and a merger. 

\subsection{Maps and radial profiles of molecular gas mass}
\label{sec:coprofiles}

We take the integrated flux CO $J=1\rightarrow 0$ maps from the public CARMA EDGE data release, convert them from their native units of Jy km/s beam$^{-1}$ to K km/s,
and then convert to H$_2$ gas mass surface densities $\Sigma_\mathrm{H_2}$ 
in units of $M_\odot$~pc$^{-2}$ by assuming 
a constant CO-to-H$_2$ conversion factor $\alpha_\mathrm{CO} = 3.1 \> M_\odot$~pc$^{-2}$~(K km/s)$^{-1}$ \citep{bolatto2013, sandstrom2013}.
The $\alpha_\mathrm{CO}$ conversion factor has 
been found to be lower by a factor of about 2 
(i.e. $\alpha_\mathrm{CO} = 1.55 \> M_\odot$~pc$^{-2}$(K km/s)$^{-1}$) in
the central kpc of nearby galaxies \citep{sandstrom2013}. 
For simplicity, we adopt $\alpha_\mathrm{CO}= 3.1 \> M_\odot$~pc$^{-2}$~(K km/s)$^{-1}$ but we do consider the impact of the 
central $\alpha_\mathrm{CO}$ on our results in later sections.
Radial profiles of $\Sigma_\mathrm{H_2}$ for our sample are computed 
in the same way as in the previous section. 
Pixels without CO detections are set to zero in the averages. 
As a result, $\Sigma_\mathrm{H_2}$ is slightly underestimated at large radii where the fraction of detected pixels is small. However,
the fraction of missing flux in the CO maps is small, so this is a good approximation to the true radial profiles \citep{bolatto2017}.
Maps and radial profiles of $\Sigma_\mathrm{H_2}$ for our example galaxies are shown in the right-most column in Figure~\ref{fig:mapsandprofiles1}.

\section{Central star formation and the link to gas concentration}
\label{sec:obsresults}

\begin{table*}	
\centering    \caption{Mean molecular gas concentrations and upturn/turnover strengths for our full sample}	\label{tab:table1}	\begin{tabular}{lcrrrr}\hline 
Category & $N$ $^\mathrm{a}$ & $\log c_\mathrm{mol}$ $^\mathrm{b}$ & $\Delta \log \mathrm{EW}(\mathrm{H}\alpha)$ $^\mathrm{c}$ & $\Delta \mathrm{EW}(\mathrm{H}\delta_\mathrm{A})$ $^\mathrm{d}$ & $\Delta\mathrm{D_n(4000)}$ $^\mathrm{e}$ \\
& & \multicolumn{1}{c}{(dex)} & \multicolumn{1}{c}{(dex)} & \multicolumn{1}{c}{(\AA)} & \\
\hline 
\multicolumn{6}{c}{Barred}\\
\hline 
All & 17 & $0.19\pm0.04$ & $0.75\pm0.12$ & $0.88\pm0.26$ & $0.20\pm0.07$ \\ 
Upturn & 13 & $0.23\pm0.05$ & $0.92\pm0.12$ & $1.13\pm0.30$ & $0.26\pm0.08$ \\ 
No upturn & 4 & $0.05\pm0.04$ & $0.17\pm0.12$ & $0.08\pm0.10$ & $0.01\pm0.10$ \\ 
\hline 
\multicolumn{6}{c}{Unbarred}\\
\hline 
All & 24 & $0.17\pm0.02$ & $0.06\pm0.03$ & $0.08\pm0.04$ & $-0.00\pm0.03$ \\ 
Upturn & 0 & $ \cdots$ & $ \cdots$ & $ \cdots$ & $ \cdots$ \\ 
No upturn & 24 & $0.17\pm0.02$ & $0.06\pm0.03$ & $0.08\pm0.04$ & $-0.00\pm0.03$ \\ 
\hline 
\multicolumn{6}{c}{Merger/pair}\\
\hline 
All & 17 & $0.27\pm0.06$ & $0.32\pm0.06$ & $0.18\pm0.08$ & $0.13\pm0.11$ \\ 
Upturn & 6 & $0.32\pm0.10$ & $0.58\pm0.06$ & $0.41\pm0.16$ & $0.14\pm0.16$ \\ 
No upturn & 11 & $0.23\pm0.07$ & $0.18\pm0.05$ & $0.04\pm0.07$ & $0.13\pm0.15$ \\ 
\hline 
\multicolumn{6}{l}{$^\mathrm{a}$ Number of galaxies in category.}\\
\multicolumn{6}{l}{$^\mathrm{b}$ Mean and uncertainty on the mean of the molecular gas concentration index (Eq.~\ref{eq:cmol}).}\\
\multicolumn{6}{l}{$^\mathrm{c}$ Mean and uncertainty on the mean of the $\log$\ewhae\ upturn strength (Eq.~\ref{eq:delta}).}\\
\multicolumn{6}{l}{$^\mathrm{d}$ Mean and uncertainty on the mean of the \ewhda\ upturn strength (Eq.~\ref{eq:delta}).}\\
\multicolumn{6}{l}{$^\mathrm{e}$ Mean and uncertainty on the mean of the \dindex\ turnover strength (Eq.~\ref{eq:delta_d4000}).}\\
\end{tabular}\end{table*}

\subsection{Recent central star formation enhancement}\label{sec:turnoverid}

Our identification and measurement of centrally elevated star formation 
is similar to the procedure described in \citet{lin2017a}.
First we calculate radial profiles of log\ewhae, \ewhda, \dindex, and
$\log\Sigma_\mathrm{SFR}$, and plot them next to the \textit{gri} 
composite image from SDSS.
We then inspect each profile in the inner region of each galaxy (inside of 
the spiral arms), and judge whether
or not the central region shows an \textit{upturn} (for $\log$\ewhae, \ewhda\ and 
$\log\Sigma_\mathrm{SFR}$) or a \textit{drop} (for \dindex) 
in the slope of the profile in the innermost region. 
The inner region corresponds to a by-eye estimate of the 
radius of the transition between bulge and disk. Note that 
\citet{lin2017a} had performed photometric decomposition on the optical images,
allowing for a precise measurement of these radii, but we do not have that information 
for all galaxies in our sample so we estimate the inner region by eye. The by-eye estimate of the inner radius is not used in 
the following analysis.
If such an upturn or turnover is identified, we mark by eye the radius 
at which it occurs.

Next, for each galaxy, and for a given star formation history indicator, 
we fit a line to the radial profile between $r=r_t$ and 
$r=r_t+5$\arcsec, where $r_t$ is the turn-up/turnover radius
determined by eye in the first step. 
This is what was done in \citet{lin2017a}. 
Those authors found that this radial range provided a sufficient fit to the
general profile in the inner region without being contaminated by spiral arms or 
the transition region.
For radial profiles which are not 
visually classified as having an upturn/turnover, we still perform the 
linear fitting, but we set $r_t$ to half of the point-spread 
function (PSF) size of CALIFA
\citep[roughly 1.25\arcsec; ][]{walcher2014}. 
In Figure~\ref{fig:mapsandprofiles1}, the linear fits 
are plotted as green solid lines for the four example galaxies. 
In each panel the radial range used for the linear fitting
is indicated by the two vertical, dotted lines.
By our definition, the value of  $\log$\ewhae~or \ewhda~at the center 
may be lower than where the fitting is performed 
(e.g. the log\ewhae~profiles of NGC4210 in Fig.~\ref{fig:mapsandprofiles1}), 
or it may be greater (as in the log\ewhae~profiles of NGC6186 and NGC5218 
in Fig.~\ref{fig:mapsandprofiles1}). In either of these scenarios, the 
value of the SFH indicator in the center is greater than expected 
from extrapolating the linear fit to the center.

The upturn/turnover strength of each galaxy, for each of 
the star formation history indicators, is then quantified
by the difference between the observed and extrapolated value 
in the central region, as measured above. Specifically, for 
$\log$\ewhae, \ewhda\ and $\log\Sigma_\mathrm{SFR}$, the upturn strength 
is defined as
\begin{equation}
\label{eq:delta}
   \Delta Y \equiv Y(r=0)-Y_\mathrm{extrap}(r=0),
\end{equation}
where $Y(r=0)$ 
is the value of $Y$ 
in the central radial bin, and 
$Y_\mathrm{extrap}(r=0)$ 
is the best-fit line extrapolated to $r=0$.
For \dindex, the turnover strength is defined as
\begin{equation}
\label{eq:delta_d4000}
    \Delta D_n(4000)\equiv D_n(4000)_\mathrm{extrap}(r=0)-D_n(4000)(r=0).
\end{equation}
Note that larger values of $\Delta D_n(4000)$ ($\Delta Y$) correspond to stronger turnovers (upturns).
The upturn and turnover strengths 
of each galaxy in our sample are listed in Table~\ref{tab:tableA2}.

In Figure~\ref{fig:obs_extrap_scatter}, which shows the central observed value of
each star formation history indicator compared to the value of the 
line extrapolated to the center for all the galaxies in our sample. 
Barred, unbarred and merging/paired galaxies are plotted as blue 
triangles, green circles and red squares, respectively.
The 1:1 relation represents no difference between the observed and extrapolated
values.
We divide our galaxies into two sets (those with or without an
upturn in $\log$\ewhae) by comparing the observed value of the 
$\log$\ewhae\ profile at $r=0$ with the value of the fitting line
extrapolated to $r=0$.
A galaxy is classified as having an upturn if it lies above the 1:1
relation (the solid line) plus the $1\sigma$ scatter (the dotted
lines) on the left panel of Figure~\ref{fig:obs_extrap_scatter}.
The scatter is the standard deviation of all points with respect to the 1:1 line.
We note that turnover/non-turnover galaxies in \citet{lin2017a}
were divided in the same way but using \dindex\
rather than $\log$\ewhae. 

\begin{figure*}
\centering
\includegraphics[width=\textwidth]{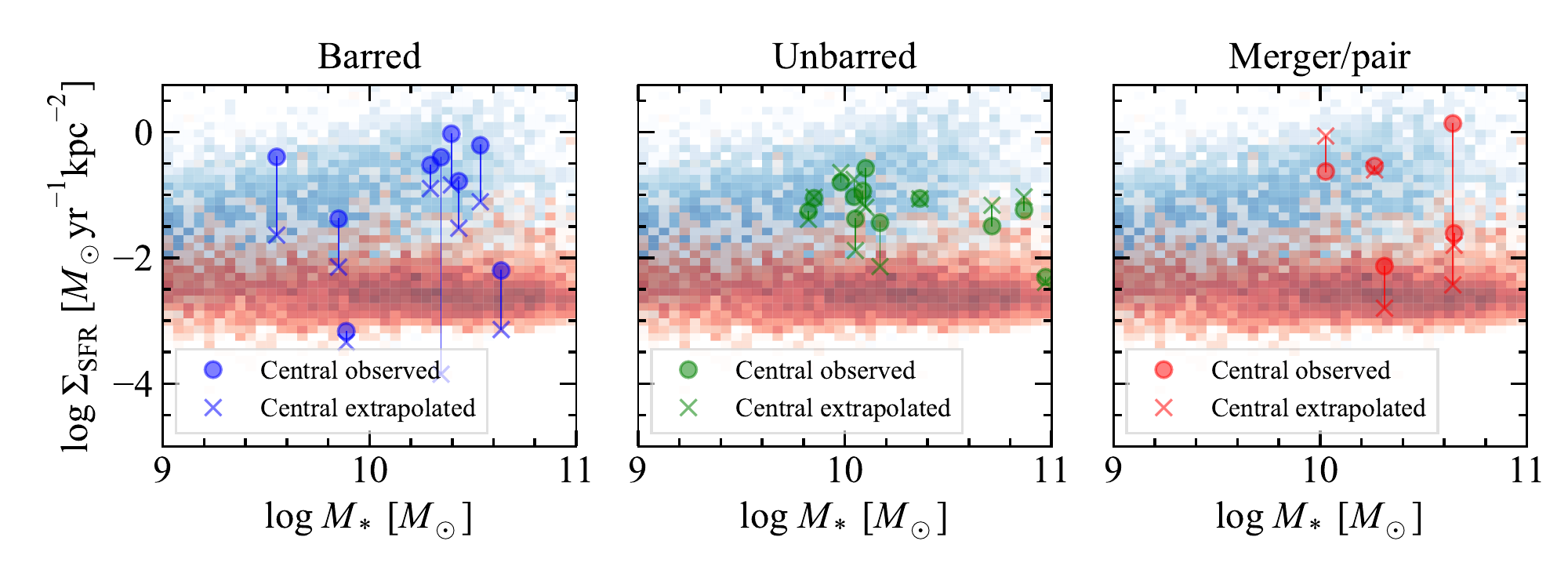} 
\caption{Observed central average (in the inner 3\arcsec) of log $\Sigma_\mathrm{SFR}$ (circles) and the linear fit extrapolated to the center (crosses) as a function of stellar mass in the SDSS fiber (provided in CALIFA DR3).  The galaxies shown in this plot are a subset of our reduced sample (\S\ref{sec:sample}) for which we can perform linear fitting to the radial profiles of $\Sigma_\mathrm{SFR}$.
The blue background population are galaxies lying within 0.5 dex of the star-forming main sequence \citep[SFMS; ][]{catinella2018}, and the red background population are galaxies more than 0.7 dex below the SFMS. We have converted the SDSS fiber SFRs to surface densities by dividing by the physical area of the fiber in kpc$^2$. This figure shows that bars are linked to increases in the central star formation rate surface density.}
\label{fig:sfrdens_vs_logmstar}
\end{figure*}

Figure~\ref{fig:tstrength_hists} shows histograms of the upturn 
and turnover strengths of our sample, for three of the star formation
history indicators:  $\Delta\log$\ewhae, $\Delta$\ewhda, and $\Delta$\dindex. 
Results are shown for the 
barred (red), unbarred (green) and paired/merging (red) galaxies 
separately in panels from left to right. 
The separating value of $\Delta\log$\ewhae~$=0.410$ dex, 
which is determined from the scatter of the points about the 1:1 
relation in Fig.~\ref{fig:obs_extrap_scatter}, is shown as the vertical dashed 
line in the top row of Fig.~\ref{fig:tstrength_hists}.
In the second and third rows of the same figure, 
we show the distributions of our sample in $\log$\ewhda\ and \dindex, highlighting the $\log$\ewhae-upturn galaxies 
as hatched histograms.
The mean and uncertainty on the mean upturn strengths from each SFH indicator 
are shown in Table~\ref{tab:table1}.

Both Figure~\ref{fig:obs_extrap_scatter} and
Figure~\ref{fig:tstrength_hists} show that none of the unbarred galaxies 
in our sample are classified as having a $\log$\ewhae~upturn. The majority 
of the upturn galaxies are barred (13/19), followed by mergers and 
pairs (6/19). On the other hand, not all of the barred or paired/merging
galaxies have upturns. This result suggests that either a bar or 
galaxy-galaxy interactions/mergers is necessary, but neither alone 
is sufficient for the central upturn to occur.  
In agreement with \citet{lin2017a}, we find that most galaxies 
classified as having a central upturn in $\log$\ewhae\
also have a relatively strong \ewhda\ upturn and \dindex\ turnover. 
These results suggest an enhancement 
in both the recent and ongoing star formation 
at the center of the upturn galaxies. 

In Figure~\ref{fig:sfrdens_vs_logmstar} we compare the central-observed
and central-extrapolated $\log\Sigma_\mathrm{SFR}$ as a function
of global stellar mass for barred (left panel), unbarred (middle 
panel) and merger/pair (right panel) galaxies. 
The galaxies shown in this plot are a subset of our reduced sample (\S\ref{sec:sample}) for which we can perform linear fitting to the radial profiles of $\Sigma_\mathrm{SFR}$.
For reference, 
in each panel we 
show the distribution of the volume-limited galaxy 
sample selected from the MPA/JHU SDSS database (see above), for 
which the SFR is measured from the SDSS 3\arcsec-fiber spectroscopy.
We find that barred galaxies generally have observed values of
$\Sigma_\mathrm{SFR}$ that are significantly higher than expected (by 
$\sim$0.5-2 dex), effectively bringing the central $\Sigma_\mathrm{SFR}$ from values that would be typical of the quiescent population or 
green valley, up into the star-forming main sequence.
For unbarred galaxies, we see little change
in the central SFR surface density, as expected. Overall, these results 
are consistent with the theoretical expectation that the central
region may be rejuvenated by star formation enhancement
driven by a bar. The merger/pair galaxies appear to either have no enhancement or centrally suppressed star formation, however, this is not representative of merger/pair galaxies as a whole due to the selection cuts.

\begin{figure*}
\centering
\includegraphics[width=0.9\textwidth]{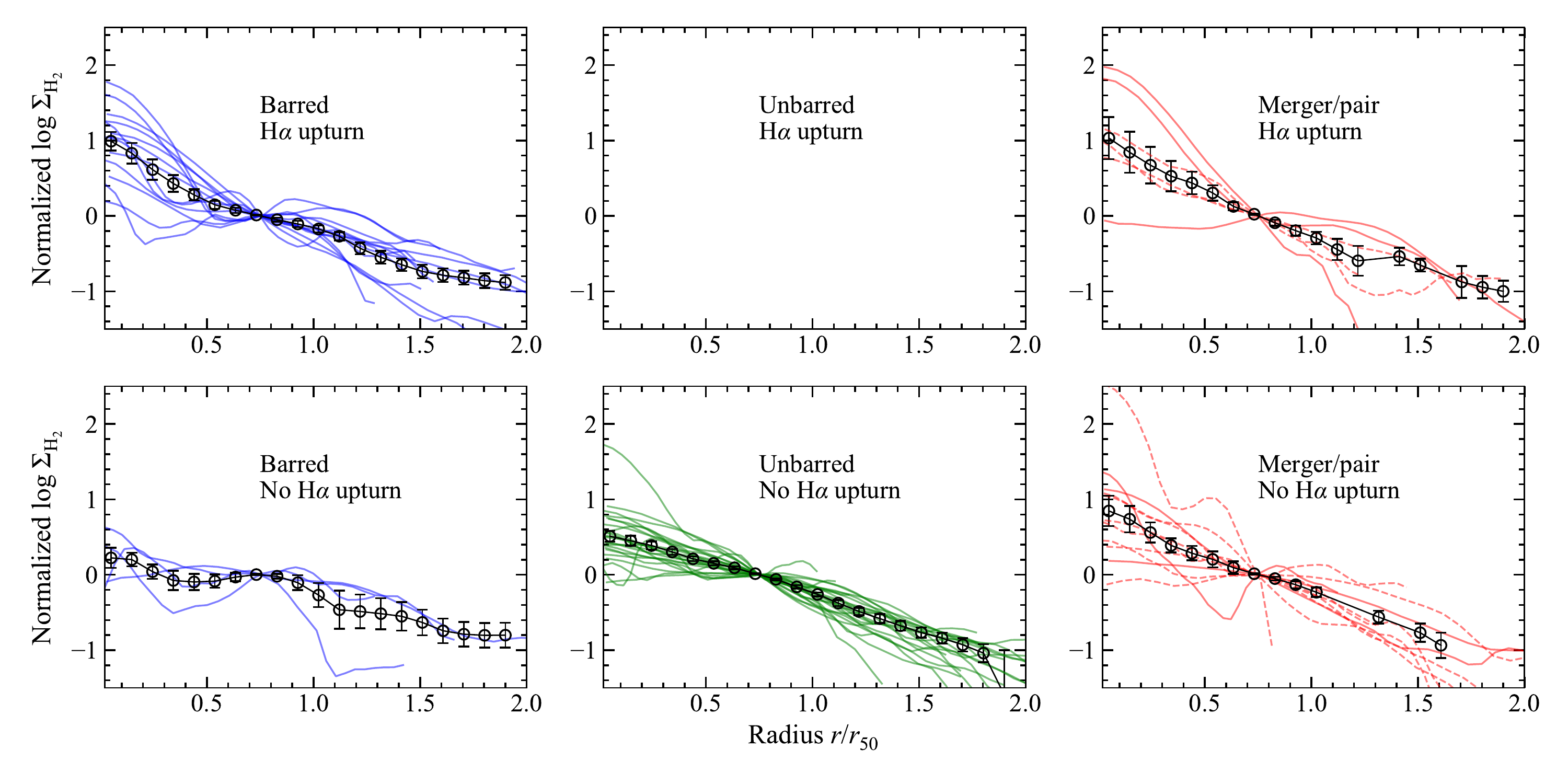} 
\caption{Azimuthally-averaged radial profiles of H$_2$ mass surface density for all galaxies in our sample. Each profile of 
$\log\Sigma_\mathrm{H_2}$ has been normalized to the value 
at $0.75r_{50}$.
In the top row, we show galaxies 
that have been identified as having a central upturn in $\log$\ewhae~(see
Sec.~\ref{sec:turnoverid}); and in the bottom row are galaxies with no central upturn. 
Note that the upper middle plot is empty because there are no unbarred galaxies in our
sample which show an $\log$\ewhae~upturn.
In each panel, the black circles and error bars are the mean and error in the mean of the profiles in each radial bin. 
The black profiles show that barred galaxies, mergers and pairs with $\log$\ewhae~upturns have 
centrally-peaked profiles with a steeper slope at $r\lesssim 0.5r_{50}$, while non-upturn barred and unbarred galaxies tend to have flatter profiles at these radii.
In the right column, barred merger/pair galaxies are shown as solid lines, while unbarred/irregular merger/pairs are shown as dashed lines.
}
\label{fig:coprofiles}
\end{figure*}

\subsection{Recent central star formation vs. molecular gas concentration}
\label{sec:gasconcentration}

Figure~\ref{fig:coprofiles} displays the H$_2$ gas
mass surface density profiles $\Sigma_\mathrm{H_2}(r)$ for 
barred (left),
unbarred (middle) and merger/pair galaxies (right), and for the subsets
of upturn galaxies (upper panels) and non-upturn 
galaxies (lower panels), as classified above. 
In this figure we have normalized each profile
to the value of $\log\Sigma_\mathrm{H_2}$ at $r=0.75r_{50}$, 
and have scaled the radius $r$ by $r_{50}$. In each panel 
we also show the mean profile and the standard deviation 
around the mean. 

The upturn galaxies
on average show a centrally-peaked
molecular gas profiles. The peak value of the normalized gas profiles relative to $0.75r_{50}$ is 
$\sim1$ dex for upturn galaxies and all merger/pairs, vs. $\sim0.5$ dex for unbarred galaxies.
Almost all of the non-upturn barred or unbarred galaxies have a 
molecular gas profile without a central peak.
Also, some 
galaxies show unusual profiles, deviating to 
varying degrees from the average profile of their type. 
These outlier galaxies are interesting targets for 
individual follow-up work.

We measure a molecular gas concentration index for each of 
our galaxies, defined by
\begin{equation}\label{eq:cmol}
    c_\mathrm{mol} \equiv \frac{r_\mathrm{50}}{r_\mathrm{50,mol}},
\end{equation}
where $r_\mathrm{50}$ is the half-light radius in the SDSS 
$r$-band, obtained from the NSA, and $r_\mathrm{50,mol}$ is the 
radius enclosing half of the total molecular gas mass. 
This definition of
concentration is similar to the optical concentration index
which is commonly used in the SDSS-based studies and defined
as $c_{r}\equiv r_{90}/r_{50}$, where $r_{90}$ is the radius 
enclosing 90\% of the \textit{r}-band light.
A larger value of $c_\mathrm{mol}$ 
indicates a higher central concentration of gas mass. 

The molecular gas half-mass radius was used in early 
single-dish surveys such as \citet{young1995},
and was estimated for many but not all (38/58) of the galaxies 
in our sample by the EDGE-CALIFA team \citep{bolatto2017}.
For completeness, we redo the measurements for all 58 galaxies in our sample.
We measure $r_\mathrm{50,mol}$ by computing the cumulative 
molecular gas mass radial profile (in linear units, not logarithmic), 
and determine the radius at which 
the enclosed mass equals half of the total H$_2$ mass.
We have adopted two different methods to measure the total H$_2$ mass.
In the first method, which is our fiducial method, we ignore non-detections
and estimate the total H$_2$ mass by the sum of the detected pixels.
In the second method, we include non-detections as 1$\sigma$ upper-limits, 
where $\sigma$ is the RMS noise in each pixel, obtained from the 
unmasked CARMA EDGE integrated flux maps. In this case the 
H$_2$ mass in each radial bin is the sum of the 
detections and non-detections, unless the 
fraction of detected pixels in that bin is less than 0.05, as was 
done in \citet{mok2017} when measuring the radial profiles of H$_2$ for galaxies in the NGLS \citep{wilson2012}. In both methods, the enclosed mass 
as a function of radius is given by the integral of the 
radial profile of $M_\mathrm{H_2}$, and
is used to determine the half-mass radius. 
We find that the two methods lead to almost identical measurements,
indicating that non-detection pixels contribute little to the
total H$_2$ mass.

All values of $c_\mathrm{mol}$ are computed using the first 
method.
Two of the galaxies in the merger/pair 
category have $r_\mathrm{50,mol}$ that is $<2\times$ 
the CARMA resolution, so we have quoted their 
concentrations as lower limits. The rest of our galaxies have 
$r_\mathrm{50,mol}$ which is significantly larger than the 
CARMA resolution.
The bottom panel of Figure~\ref{fig:tstrength_hists} shows 
the distribution of molecular gas concentrations for our sample 
as a whole and for the subsamples of barred, unbarred and 
merger/paired galaxies separately. 
The molecular gas concentrations of the galaxies 
in our sample are also listed in Table~\ref{tab:tableA2}.

\begin{figure*}
\centering
\includegraphics[width=0.9\textwidth]{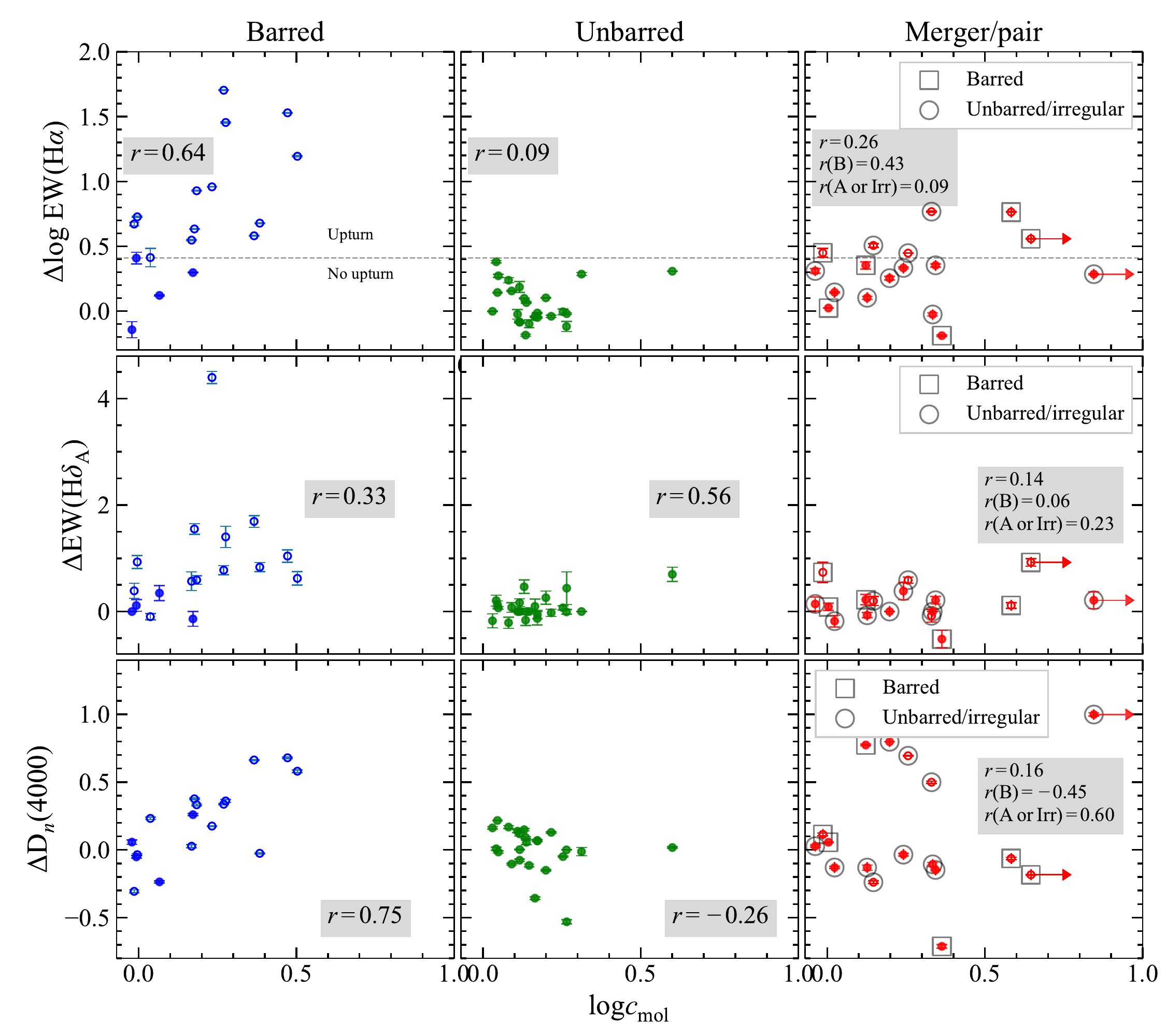} 
\caption{
$\log$ \ewhae~upturn strength \textit{(top row)}, \ewhda~upturn strength \textit{(middle row)}, and \dindex~turnover strength \textit{(bottom row)} 
as a function of molecular gas concentration. The open points
are those which are classified as having a central $\log$\ewhae~upturn (see Sec.~\ref{sec:turnoverid}),
while the filled symbols have no upturn. The error bars are the uncertainties on the measured 
central value of the SFH indicator. The value of $\Delta\log$\ewhae~which 
divides these two categories is shown by the horizontal line in the top row. The Pearson correlation coefficient $r$ is shown for each panel.
This figure shows our main result: galaxies which show centrally enhanced 
recent star formation are either barred or in a merger/pair, and on average have 
higher gas concentrations. 
}
\label{fig:strengthvsco}
\end{figure*}

 We have also compared our 
$r_\mathrm{50,mol}$ with those of \citet{bolatto2017},
and find the two to agree well, with no systematic differences.
The mean absolute differences in $r_\mathrm{50,mol}$ are 1.6\arcsec\ and 1.8\arcsec\ for the two 
methods adopted in our case, comparable to the 1\arcsec\ pixel size.
In \citet{sheth2005} 
the total H$_2$ masses were obtained from single-dish CO 
measurements, while the nuclear H$_2$ mass were obtained from 
spatially resolved CO maps. 
We do not have single-dish CO data, but, as discussed 
in \citet{bolatto2017},
the total flux in the CARMA EDGE maps matches
well with expectations 
based on single-dish CO calibrations from \citet{saintonge2011a}.
We have also investigated the potential impact of a central 
drop in the CO-to-H$_2$ conversion factor $\alpha_\mathrm{CO}$ 
in our analysis, and find that the change in $c_\mathrm{mol}$ 
is negligible. 

In Table~\ref{tab:table1} we show the mean and error on the mean of $\log c_\mathrm{mol}$ for barred, unbarred, and merger/pair galaxies. We have calculated these quantities for all galaxies in each category, and the upturn/non-upturn galaxies separately. Galaxies with an upturn (barred or merger/pair) have significantly higher concentrations than unbarred galaxies. Barred galaxies without an upturn have significantly lower concentrations than all other categories. Interestingly, merger/pair galaxies without an upturn have relatively high gas concentrations, which are consistent on average with those with an upturn. These results show that, in order to have an upturn, it is not enough to be a merger/pair with high molecular gas concentration.

\begin{figure}
\includegraphics[width=0.45\textwidth]{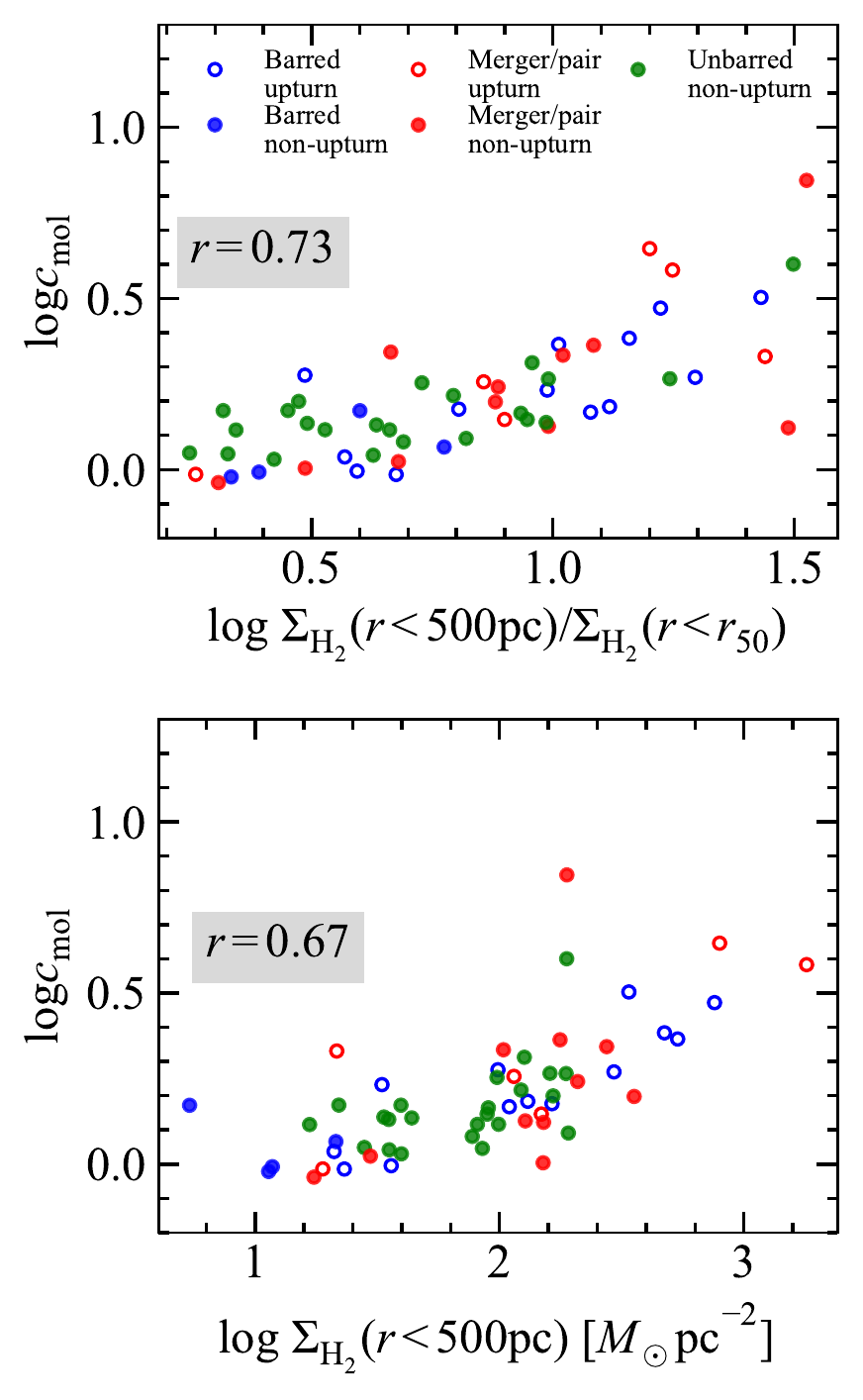} 
\caption{\textit{Top:} Correlation between molecular gas concentration and the concentration defined by \citet{sheth2005}, namely $\Sigma_\mathrm{H_2}$ calculated in the central 1 kpc divided by $\Sigma_\mathrm{H_2}$ calculated over the whole CO disk (which is approximately $r_{50}$). \textit{Bottom:} Correlation between molecular gas concentration and $\Sigma_\mathrm{H_2}$ calculated in the central 1 kpc. The Pearson correlation coefficient $r$ using all galaxies is shown in both panels. }
\label{fig:coconc_vs_sheth}
\end{figure}

In Figure~\ref{fig:strengthvsco} we examine
the correlation of central star formation enhancement with
molecular gas concentration, by 
showing the $\Delta\log$\ewhae\ (first row), 
$\Delta$\ewhda\ (second row),
and $\Delta$\dindex\ (third row) 
as a function of $\log c_{\mathrm{mol}}$
for the three main categories (barred, unbarred, and merger/pairs). 
The Pearson correlation coefficients shown in Fig.~\ref{fig:strengthvsco} show that the strongest correlations are between $\log\Delta$\ewhae\ and $\log c_\mathrm{mol}$ for barred galaxies, and between $\Delta$\dindex\ and $\log c_\mathrm{mol}$ for barred galaxies. There also appears to be some correlation between $\Delta$\ewhda\ and $\log c_\mathrm{mol}$ for unbarred galaxies. There does not appear to be a significant correlation between upturn strength and concentration in other panels.
The lack of correlation between central SF enhancement and $c_\mathrm{mol}$ in merger/pair galaxies may be suggesting that the enhancement may be episodic. It is interesting that there are some merger/pair galaxies with significant enhancements in \dindex\ but not in \ewhda\ or $\log$\ewhae\, and with relatively low gas concentrations.

Here we briefly remark on why we did not measure a concentration index like 
$c_\mathrm{mol}$ for the SFH indicators.
The gas concentration is a proxy for how much gas has been transported inwards. 
Generally, the gas radial profiles are either flat or with a central peak. This could possibly be due to the lower resolution of the CO data. The case for 
the SFH indicators is more complex -- there are cases where there is clearly a central 
peak/drop in SFH indicators, but this is sometimes a local peak or drop. 
A concentration index is sensitive 
to where most of the gas or SF is happening, and would not pick out a central peak if 
it is a local one. This problem could be 
alleviated by calculating a concentration within the inner region only, however 
this would require a by-eye estimation of the radius of the inner regions since we 
have not performed photometric decomposition.
This is why we decided to measure upturn strengths on the central region specifically.

Now we assess whether such correlations may be due to our particular definition of 
molecular gas concentration. The top panel of Figure~\ref{fig:coconc_vs_sheth} shows the comparison between our concentration parameter and an alternative definition from \citet{sheth2005}, namely $\Sigma_\mathrm{H_2}$ in the central kiloparsec divided by the total $\Sigma_\mathrm{H_2}$. In their analysis, the total $\Sigma_\mathrm{H_2}$ was obtained from single-dish CO measurements, whereas we measure it from the CARMA CO maps. Specifically, we measure the total H$_2$ mass within $r_{50}$ and divide by the area of the corresponding ellipse. The Pearson correlation coefficient between these quantities is 0.73, indicating a good correlation. Thus, if we were to use this alternative concentration in our analysis, our results would not change significantly. The lower panel of Fig.~\ref{fig:coconc_vs_sheth} shows our concentration versus $\Sigma_\mathrm{H_2}$ calculated in the central kpc (500 pc radius). Again, the correlation coefficient of 0.67 suggests a significant correlation, and our results would also not change significantly if we were to use $\Sigma_\mathrm{H_2}$ in the central kpc in place of $c_\mathrm{mol}$.

We do not find significant correlations between $c_\mathrm{mol}$ and parameters which quantify 
the global properties of a galaxy, namely global stellar mass (from the NSA; $r=0.16$), \nuvr~colour (from the NSA; $r=0.29$), optical concentration ($r_\mathrm{90}/r_\mathrm{50}$, from the NSA; $r=0.41$), H$_2$ mass fraction ($M_\mathrm{H_2}/M_*$; $r=0.33$), or \ion{H}{i} mass fraction using \ion{H}{i} measurements from the ALFALFA 100\% catalog \citep{haynes2018} 
or the HyperLeda database \citep{makarov2014} ($r=-0.23$). 
We find a slight correlation between $\log c_\mathrm{mol}$ and
H$_2$-to-\ion{H}{i} mass ratio ($r=0.48$). In a sense, this quantity 
is also a gas concentration, because it is a measure of the gas mass in a small area (H$_2$ tends to be more centrally concentrated) divided by the gas mass in a large region (\ion{H}{i} is known to have a larger spatial extent than H$_2$), so this tentative correlation is not surprising.
These results indicate that the
gas concentration is indeed caused by bars or mergers, an effect
which is independent of galaxy mass, global color or light concentration.

\begin{figure*}
\centering
\includegraphics[width=\textwidth]{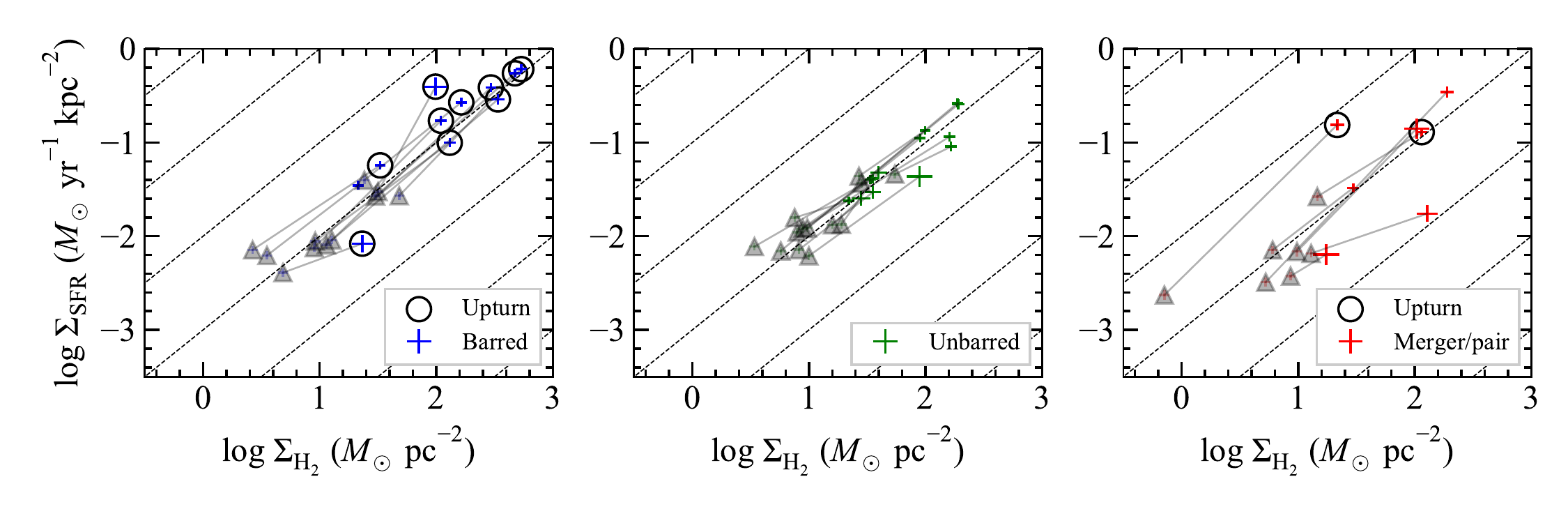} 
\caption{$\log\Sigma_\mathrm{SFR}$ vs. $\log\Sigma_\mathrm{H_2}$ in the 
central 0.5 kpc (in semi-major axis; coloured symbols) and over annuli covering $0.5$kpc $<r\leq r_{50}$ (in semi-major axis; grey triangles). These two are connected by lines for each galaxy. 
There are fewer galaxies shown here than in Fig.~\ref{fig:coprofiles} because here we are using the reduced sample, described in \S\ref{sec:sample} and \S\ref{sec:morphclass}.
Galaxies classified as having a $\log$\ewhae\ upturn are identified with black 
open circles. 
The same (Galactic) CO-to-H$_2$ conversion factor was used for all points -- if a lower $\alpha_\mathrm{CO}$ were used in the central kpc, the central points would shift leftward by 0.3 dex.
The diagonal lines indicate constant H$_2$ depletion times.
This plot shows where barred/unbarred/interacting and upturn/non-upturn
galaxies lie in terms of their \textit{absolute values} of
$\log\Sigma_\mathrm{SFR}$ and $\log\Sigma_\mathrm{H_2}$. 
It is clear that barred galaxies extend to higher $\Sigma_\mathrm{SFR}$ and 
$\Sigma_\mathrm{H_2}$ in the central region than unbarred galaxies. 
The mean and uncertainty on the mean for each parameter are shown in
Table~\ref{tab:table2}.
}
\label{fig:ks}
\end{figure*}

\begin{table*}	
\centering    \caption{Mean surface densities of H$_2$ and SFR, and depletion times for our reduced sample (described in \S\ref{sec:sample})}	\label{tab:table2}	\begin{tabular}{lrccccccc}\hline 
   Upturn? $^\mathrm{a}$  & $N$ $^\mathrm{b}$ & \multicolumn{2}{c}{$\log\Sigma_\mathrm{H_2}$ $^\mathrm{c}$} & \multicolumn{2}{c}{$\log\Sigma_\mathrm{SFR}$ $^\mathrm{d}$} & \multicolumn{2}{c}{$\log\tau_\mathrm{depl.}$ $^\mathrm{e}$} & $\log(\tau_\mathrm{cen.}/\tau_\mathrm{disk})$ $^\mathrm{f}$ \\ 
 (Y/N) & & \multicolumn{2}{c}{$(M_\odot$pc$^{-2})$} & \multicolumn{2}{c}{$(M_\odot$yr$^{-1}$kpc$^{-2})$} & \multicolumn{2}{c}{(yr)} & (dex) \\  & & Center & Disk & Center & Disk & Center & Disk & \\ 
\hline 
\multicolumn{9}{c}{Barred}\\
\hline 
 Y & 10 & $2.16\pm0.15$ & $1.12\pm0.12$ & $-0.75\pm0.18$ & $-1.89\pm0.11$ & $9.05\pm0.09$ & $9.15\pm0.06$ & $-0.10\pm0.09$ \\ 
 N & 1 & $1.33\pm0.04$ & $0.55\pm0.04$ & $-1.46\pm0.05$ & $-2.21\pm0.04$ & $8.92\pm0.07$ & $8.89\pm0.06$ & $0.03\pm0.09$ \\ 
\hline 
\multicolumn{9}{c}{Unbarred}\\
\hline 
 Y & 0 & $\cdots$ & $\cdots$ & $\cdots$ & $\cdots$ & $\cdots$ \\ 
 N & 13 & $1.84\pm0.10$ & $1.08\pm0.09$ & $-1.17\pm0.10$ & $-1.85\pm0.08$ & $9.14\pm0.04$ & $9.06\pm0.05$ & $0.07\pm0.04$ \\ 
\hline 
\multicolumn{9}{c}{Merger/pair}\\
\hline 
 Y & 2 & $1.70\pm0.36$ & $0.51\pm0.66$ & $-0.83\pm0.42$ & $-2.10\pm0.53$ & $8.68\pm0.40$ & $8.74\pm0.13$ & $-0.06\pm0.27$ \\ 
 N & 5 & $1.82\pm0.20$ & $0.91\pm0.07$ & $-0.93\pm0.32$ & $-2.28\pm0.07$ & $9.31\pm0.21$ & $9.32\pm0.07$ & $-0.02\pm0.18$ \\ 
\hline 
\multicolumn{9}{l}{Note: a Galactic $\alpha_\mathrm{CO}$ was assumed, and a 10\% calibration uncertainty is included in both $\Sigma_\mathrm{H_2}$ and $\Sigma_\mathrm{SFR}$.}\\
\multicolumn{9}{l}{Here, ``center'' refers to $r<500$ pc, and ``disk'' refers to $500$ pc $\leq r \leq r_\mathrm{50}$. }\\
\multicolumn{9}{l}{$^\mathrm{a}$ Galaxies which have an upturn in $\log$\ewhae\ or not.}\\
\multicolumn{9}{l}{$^\mathrm{b}$ Number of galaxies in category.}\\
\multicolumn{9}{l}{$^\mathrm{c}$ Mean and uncertainty on the mean of $\log\Sigma_\mathrm{H_2}$ in the center and disk.}\\
\multicolumn{9}{l}{$^\mathrm{d}$ Mean and uncertainty on the mean of $\log\Sigma_\mathrm{SFR}$ in the center and disk.}\\
\multicolumn{9}{l}{$^\mathrm{e}$ Mean and uncertainty on the mean of $\log\tau_\mathrm{depl.}$ (Eq.~\ref{eq:tdepl}) in the center and disk.}\\
\multicolumn{9}{l}{$^\mathrm{f}$ Mean and uncertainty on the mean of the center-to-disk depletion time ratio.}\\
\end{tabular}\end{table*}

\subsection{Linking central star formation enhancement with molecular gas mass profiles}
\label{sec:profiles}

For another perspective, rather than comparing the \textit{relative} enhancement in star formation history indicators between barred/unbarred and merger/pair categories, here we compare the absolute values of $\log\Sigma_\mathrm{SFR}$ and $\log\Sigma_\mathrm{H_2}$ in these categories. For this analysis we have used the reduced sample (31 galaxies) described in \S\ref{sec:sample}. Figure~\ref{fig:ks} shows $\log\Sigma_\mathrm{SFR}$ and $\log\Sigma_\mathrm{H_2}$ for barred (left), unbarred (middle), and merger/pair galaxies (right). For each galaxy we show the means and uncertainties of these surface densities, calculated in the central kpc (500 pc radius, shown as the coloured symbols) and in the disk ($r\leq r_{50}$, triangles). The uncertainties include measurement errors as well as a 10\% calibration uncertainty in each quantity. A Galactic $\alpha_\mathrm{CO}$ was used for all points. 

We use these measurements to compute molecular gas depletion times 
\begin{equation}\label{eq:tdepl}
    \tau_\mathrm{depl.} \equiv \frac{\Sigma_\mathrm{mol}}{\Sigma_\mathrm{SFR}},
\end{equation}
where $\Sigma_\mathrm{mol}=1.36\Sigma_\mathrm{H_2}$ (where the factor of 1.36 accounts for the presence of helium).
We have measured the mean and uncertainty on the mean of $\log\Sigma_\mathrm{H_2}$, $\log\Sigma_\mathrm{SFR}$, $\log\tau_\mathrm{depl.}$, and $\log(\tau_\mathrm{cen.}/\tau_\mathrm{disk})$, all of which are shown in Table~\ref{tab:table2}. The measurements for each galaxy are given in Table~\ref{tab:tableA2}. Table~\ref{tab:table2} shows that barred galaxies with an upturn have a slight drop in depletion time in their centers compared to their disks, while unbarred galaxies have a slight central rise in depletion time. Although $\alpha_\mathrm{CO}$ in the central kpc may be lower by a factor of 2 \citep{sandstrom2013}, this would shift all $\log(\tau_\mathrm{cen.}/\tau_\mathrm{disk})$ lower by 0.3 dex, without changing their uncertainties. This would turn the central rise in $\tau_\mathrm{depl.}$ for unbarred galaxies into a slight drop, but would turn the drop for barred galaxies into an even larger drop. However, the \textit{relative} value of these quantities between the barred, unbarred and merger/pair categories should not be affected by such variations in $\alpha_\mathrm{CO}$. 

We would like to emphasize that due to the small number statistics here, we do not interpret these findings as strong evidence. The best statistical comparison we can make is between barred upturn galaxies and unbarred galaxies -- we find $\log(\tau_\mathrm{cen.}/\tau_\mathrm{disk})$ is lower in barred upturn galaxies than unbarred galaxies by $0.17\pm 0.10$ dex. 
Merger/pair galaxies do not show a statistically significant increase or decrease in depletion times in their centers on average, but from Fig.~\ref{fig:ks}, one can see that some of these galaxies have $\log(\tau_\mathrm{cen.}/\tau_\mathrm{disk}) \leq 0$, while others have $\log(\tau_\mathrm{cen.}/\tau_\mathrm{disk}) > 0$.

These results are similar to those of \citet{utomo2017}, who found $\log(\tau_\mathrm{cen.}/\tau_\mathrm{disk})$ for barred and interacting galaxies to be $-0.22\pm 0.28$ dex and $-0.42\pm 0.51$ dex respectively. They found this ratio to be $-0.03\pm 0.35$ dex for unbarred galaxies. Those authors used a slightly different definition of ``center'' and ``disk,'' which may be the reason for an offset between our ratios and theirs, however the relative difference between barred, unbarred and interacting is similar to what we find.


\section{Comparison to an $N$-body simulation}
\label{sec:simresults}

\subsection{Observations}

To understand our results better, our observational results are compared to 
high-resolution 
$N$-body hydrodynamic simulations. Given the large
diversity of properties of our observed 
barred galaxies (properties of the disc, the bulge(s), the bar as well
as of the spirals and rings that the latter may drive, etc.) it is
clearly well beyond the scope of this paper to attempt truly quantitative 
comparisons. We will thus restrain ourselves here to {\it qualitative}
comparisons and to generic properties. 

We selected NGC 5000 (Fig.~\ref{fig:simmap}) as a random galaxy from our sample, using only the rough
morphological constraints that the chosen galaxy should have a
strong bar, a clear inner ring, and that it be viewed nearly face-on. 
We then calculated the mass-weighted average age of the 
stellar population for each IFU spaxel
by fitting the corresponding spectrum with
a linear combination of simple stellar populations (SSP) of different
ages. Then, using simple azimuthal averages, we obtained the
corresponding radial age profile (Fig.~\ref{fig:simprofile}). For such fits it is customary to include populations considerably 
older than the Universe, e.g. up to 18 Gyr, or even more \citep[][ etc.]{gonzalez-delgado2015, scott2017, ge2018}
 in
order to ensure completeness of templates. Since we
want to compare with $N$-body simulations following the formation of
discs and of their bars, we limited the age range of our
SSPs to 13 Gyr. 

\subsection{Simulations}
One of us (EA) has run a number of high resolution $N$-body
hydro-dynamic simulations, including star formation, feedback, and
cooling. They follow the formation of a disc and of its components,
such as the bar, B/P bulge, lens, spirals, rings etc. 
We give a very short summary of the simulations here; 
readers who want more information should refer to \citet[][ hereafter A16]{athanassoula2016} 
and \citet{rodionov2017}.

The simulations were run using a code based on Gadget3 
\citep{springel2002, springel2003, springel2005}, with only a few
modifications (A16 and references therein).
The initial conditions and the way they were built are also described
in that paper. This simulation survey includes galaxies
formed and evolved in isolation, as well as galaxies formed from the
merging of an isolated pair of protogalaxies, with the merging
occurring early on in the run. Compared to most other dynamical
simulations, where the disc is already in place and fully developed in
the initial conditions of the simulation, they have the
big advantage of not starting off with a pre-fabricated disc
component, which might unwittingly introduce a bias. Indeed, they
start off as single spheroid, composed of dark matter and hot gas, or
as a pair of such spheroids. 
The dark matter is described by 2.2 million collisionless
particles and the hot gas by 1.25 million smoothed particle hydrodynamics
(SPH) particles. The linear resolution is equal to 50 pc for the dark
matter and 25 pc for the baryonic particles.

During the evolution, the gas cools by radiation 
and accretes onto a plane perpendicular to the initial halo spin axis,
thus forming a gaseous disc which, via star formation, becomes partly
stellar. This disc forms inside-out, growing with time in radial extent
while its gas-to-total-baryonic mass ratio decreases (A16), in good
agreement with observations of galaxies at higher and lower
redshifts. A stellar bar starts forming and evolves at the same time
as the disc.  

The various simulations we had at our disposal, as well as most other
simulations in the literature, were run for a time span of at
most 10 Gyr. Since our goal is to compare the radial age
profile of our simulations with that derived from SSP fits to the
CALIFA data of NGC 5000, we continued our  
run to 13 Gyr. 
We will hereafter call $t=13$ Gyr the final
time of the simulation. 

\subsection{Morphology}

\begin{figure*}
\centering
\includegraphics[width=0.3\textwidth]{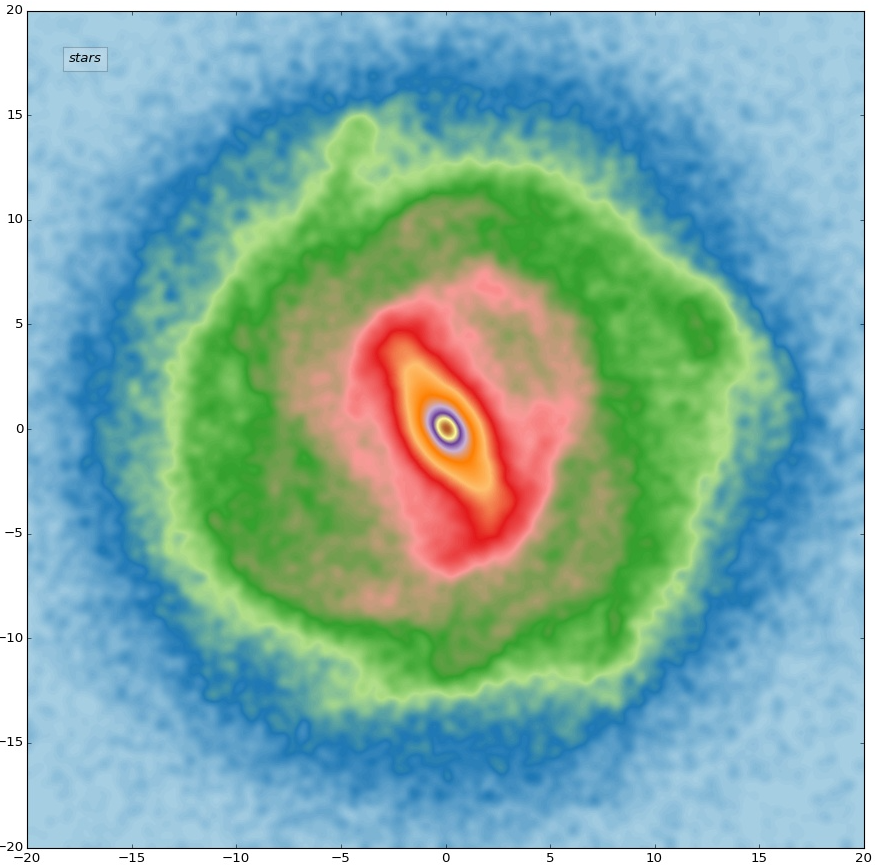}\includegraphics[width=0.3\textwidth]{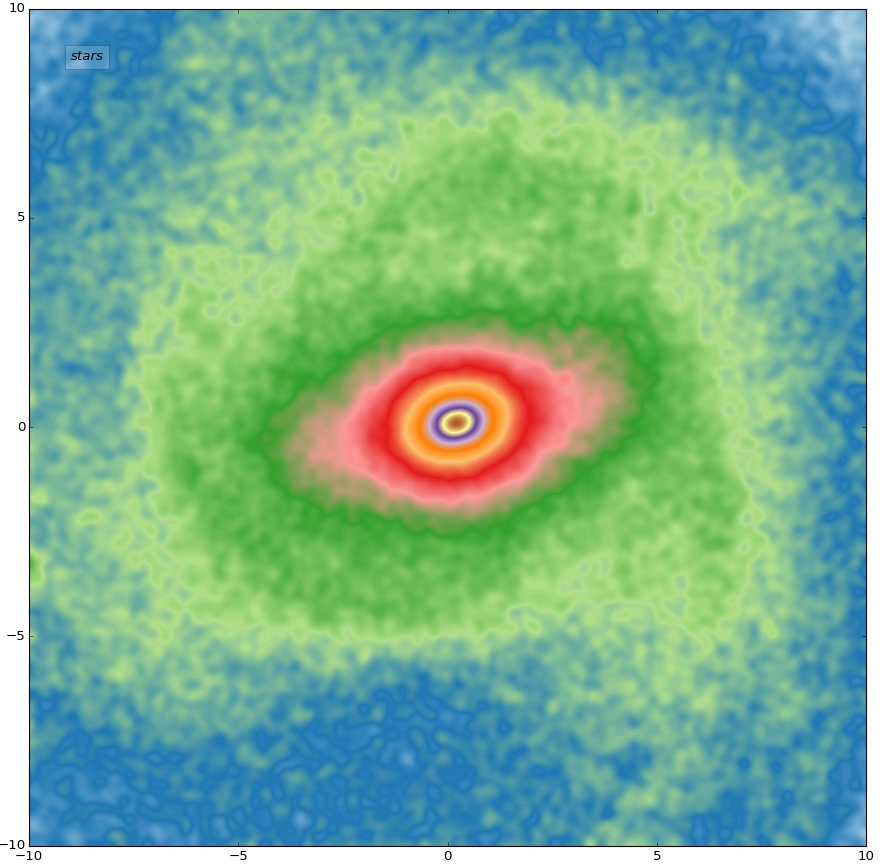}\includegraphics[width=0.3\textwidth]{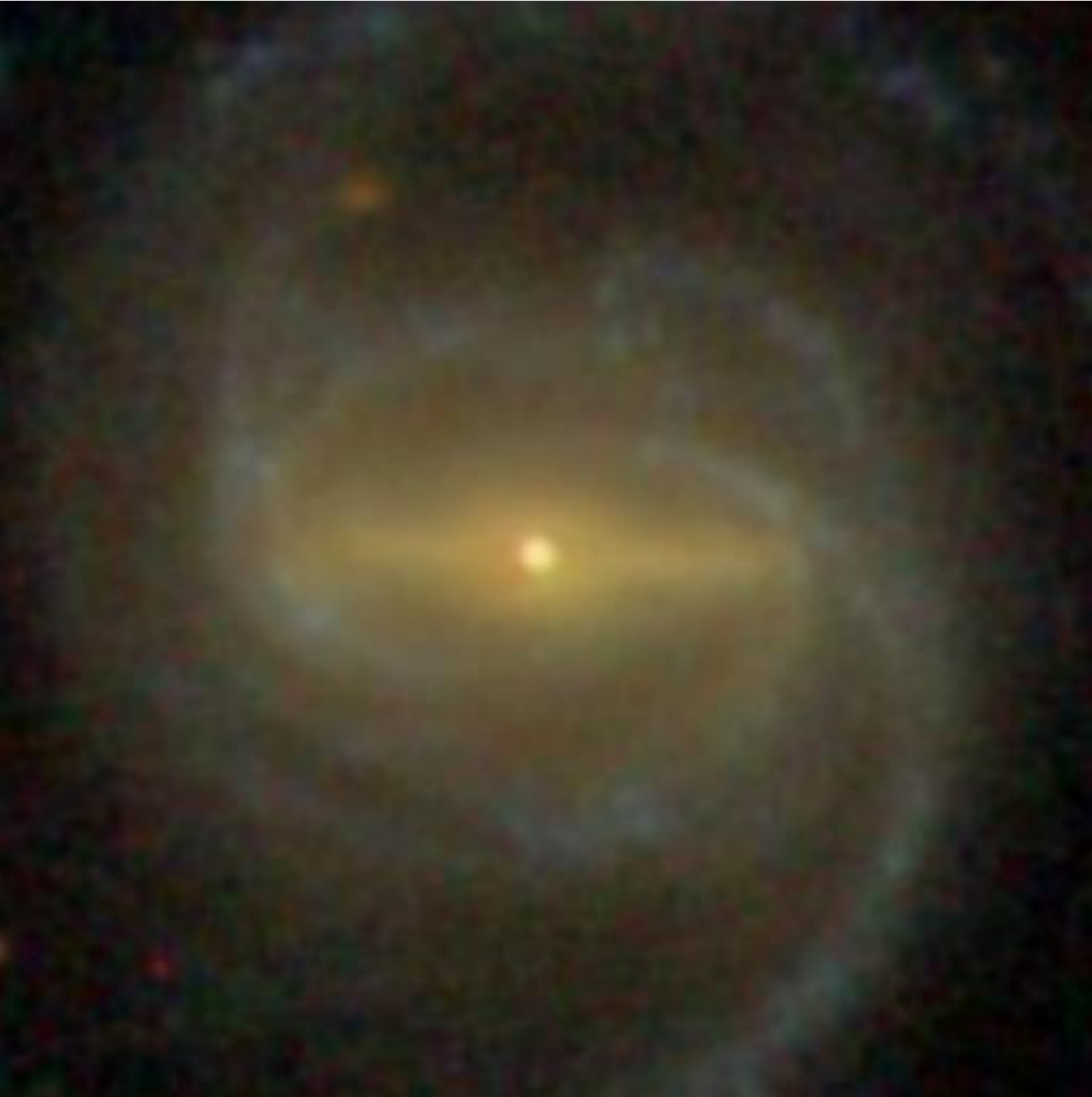}
\caption{The left and middle panels show stellar maps of a snapshot of
our simulated galaxy at times 9 and 13 Gyr, respectively. The image
dimensions are 40 kpc $\times$ 40 kpc for the former and 20 $\times$
20 for the latter. The right panel shows the
optical \textit{gri} image (29.3 kpc $\times$ 29.3 kpc) of NGC 5000, a
galaxy from our sample. These images, and particularly their
qualitative similarities, are discussed in Sect. 4.3.
}
\label{fig:simmap}
\end{figure*}

None of the simulations at our disposal reproduce well all the morphological features of NGC 5000. Many had a partial success, i.e. reproduced well some features
but never all. This is no
surprise, given the very large variety of disc galaxy properties. Thus
the probability of having a simulation which will reproduce a given
observed galaxy is very small, unless we run a large number of
simulations specifically for this task,
which is well beyond the
scope of this paper. 
Fig.~\ref{fig:simprofile} shows that  
the maximum value of the radial age profile for NGC
5000 is around 12.5 Gyr. To reach such ages, we thus need to constrain
our comparisons to times as near the final time as possible and
certainly above 12.5 Gyr. This sets very strong restrictions to our
choice. 

We chose a snapshot in the same way 
as we had chosen NGC 5000, i.e. a snapshot which has a bar and
an inner structure such as a lens or inner ring (Fig.~\ref{fig:simmap}).
It has a clear bar; however, compared
to the observations, the simulated bar is less thin, and thus
is presumably not as strong. 
Both the real and the
simulated bar are surrounded by a  component which is fatter in the
equatorial plane. 
In the real galaxy, however, this component is more 
ring-like, while in the simulated one it is more lens-like. This
relatively minor difference will have some influence on the shapes of
the age radial profiles. 
Outside this
structure the observations show a grand design two-armed spiral structure. 
This exists also in the stellar component of the simulation, but has a much
lower amplitude. This difference must be due to the fact that we are
looking at times as late as 13 Gyr, so that the bar and, albeit to a
lesser extent, the spirals 
have had ample time to stir up the stars in the region outside the
ring and to increase their velocity dispersion considerably, so that
no high amplitude spirals can be driven \citep[for a review, see][]{athanassoula1984}. This 
could also explain the difference of the bar shape, because
horizontally thinner bars can be found in discs with lower velocity
dispersions than thicker bars 
\citep{athanassoula1983, athanassoula2003, athanassoula2018, bekki2011, athanassoula2016, fragkoudi2017}.

As this comparison is not very satisfactory, most likely 
due to the late comparison
time, we decided to look also at earlier times. We found a much
better resemblance between the observed and simulated morphologies.
For example, at a time of 9 Gyr (left panel), we find a much
thinner bar (although still not as thin as that of NGC 5000), the
outline of a partial inner ring (rather than 
a lens-like enhancement) and much clearer outer spirals.
We thus decided to include this time also in the comparisons, even
though this excluded stars older than 9 Gyr from the comparisons,
i.e. has a worse completeness of SSP templates.    

\subsection{Age radial profile}

\begin{figure}
\centering
\includegraphics[width=0.5\textwidth]{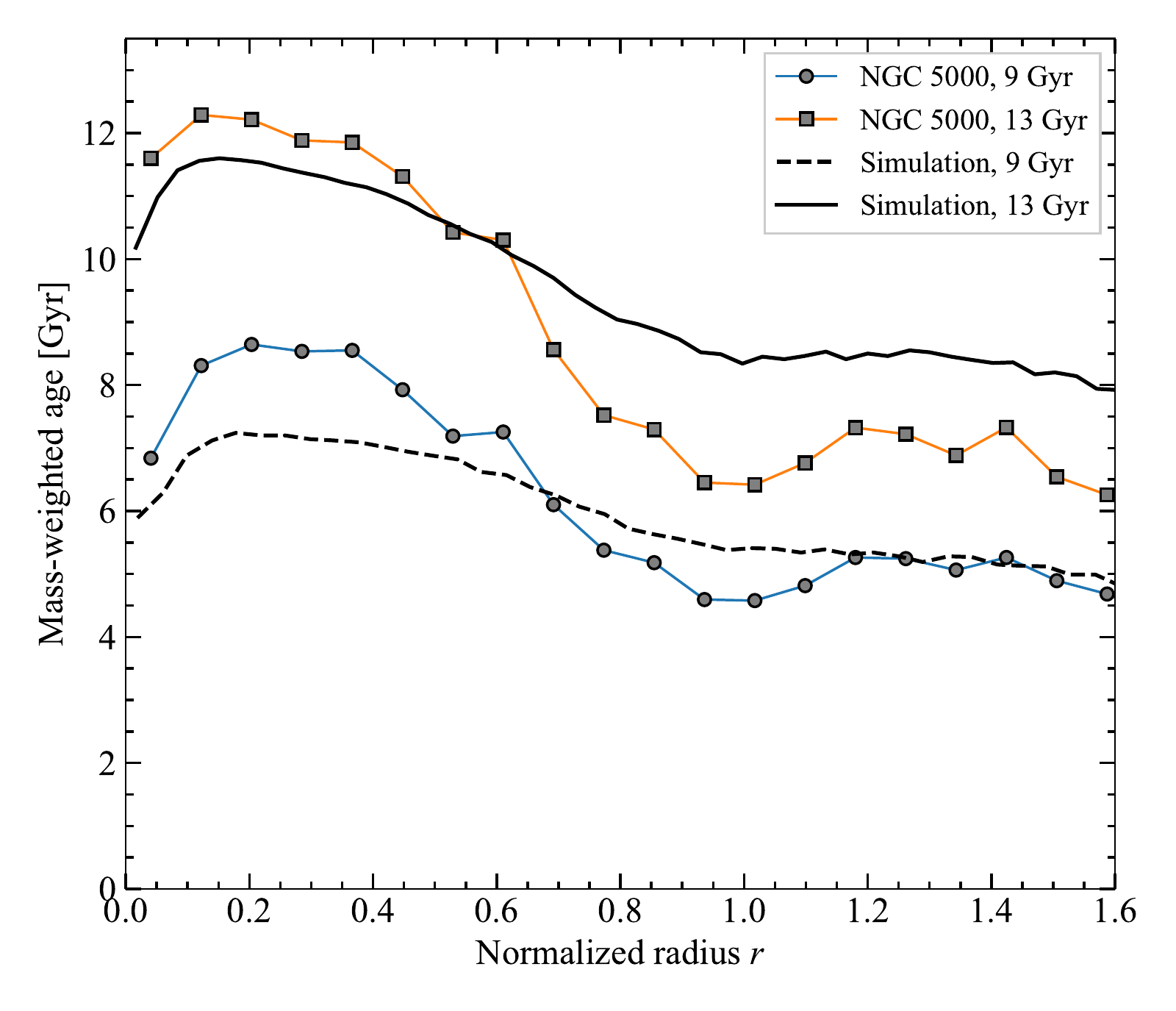}
\caption{ Comparison of the azimuthally-averaged mass-weighted stellar age 
radial profile of NGC 5000 (circle and square symbols) with that
of the simulated galaxy. SSPs with ages up to 13 Gyr (squares), or
up to 9 Gyr (circles) were used to fit the CALIFA data here. 
The radii have been normalised to the location of the deepest minimum (corresponding to the inner ring),
which is near the end of the bar (the local minimum is at a value of
1.0 on the x-axis). The two main points to note are the central drop in
age and the fact that the profile decreases after the end of the drop.
The former could be associated with the inner Lindblad
resonance of the bar, while the latter is evidence for the disc's and
the bar's inside-out formation.}
\label{fig:simprofile}
\end{figure}

In order to compare the two pairs of azimuthally-averaged radial age
profiles, we need to calibrate the radii for the data and the simulation. The best calibration is by using physically-motivated lengths, as bars can have considerably different lengths in kpc in the two cases. 
A good candidate is the radius of one of the main resonances, and the one that is
easiest to measure is the corotation or the inner ultra-harmonic
resonance. This is one of the reasons we chose a galaxy with a  clear
inner ring (corresponding to the deepest age minimum), as in that case the radius of the ring can be used as a
measuring rod. An inner ring is also easily observed as it is the
locus of intense star formation and thus can be seen as a clear minimum
in any azimuthally averaged age profile. 

We calculated the azimuthally-averaged age radial profile from the
simulation both at 9 and 13 Gyr, and we compare them in
Fig.~\ref{fig:simprofile} with two age radial profiles of NGC 5000, calculated with maximum SSP ages of 9 and 13 Gyr. Given that the simulation we
use was not specifically run to match NGC 5000, that there are
considerable morphological differences between the simulations and
observations, and that there are no free parameters to improve the
match, we could have expected a very poor fit. Our results, however,
are good qualitatively and even acceptable quantitatively. 

Over the full radial range used in Fig.~\ref{fig:simprofile}, the 
average difference between the simulation and the SSP fitted
observations is considerably smaller for 9 Gyr than for 13 Gyr (0.625
Gyr for the former, compared to 1.127 Gyr for 13 Gyr). 
Considering separately 
the part of the disc within a normalised radius of 0.6 and the part
beyond that radius, we see that the 13 Gyr model fits better in the
former (0.56 Gyr average absolute difference for 13 Gyr, compared to 1.08 for 9
Gyr), while the 9 Gyr model fits better in the region beyond the radius
of 0.6 (0.37 Gyr average absolute difference, compared to 1.43 for the
13 Gyr model). 
If we compare the values of the maximum age in the profile, which
are in all cases located in the inner part, the difference between
simulations and SSP fits is around 6\% for the 13 Gyr case, which is quite good, and
better than 20\% for the 9 Gyr case, which is acceptable given the
differences between morphologies of the simulation and NGC 5000. 
What is most interesting, however, is that we can see a few generic
features, present for both times and for the observations as well as
the simulations. Two are most important, namely that the profile has a
central dip, and that beyond that dip the age decreases with
increasing distance from the centre. 

In the central part of all four profiles there is a clear age minimum of relatively small
radial extent. The size of
this minimum shows a good agreement between observations and
simulations. With the help of the simulation used here \citep[but see also
previous simulations, e.g.][]{athanassoula1992b} it is easy to
see how this minimum is formed, as the bar exerts
torques on the gas and pushes it inwards. Thus, the gas concentrates in
a small central area, where it reaches very high
densities. \citet{athanassoula1992a} showed that the radius of this area
is set by the Inner Lindblad Resonance (ILR) radius (i.e. by the
extent of the x2 orbits). This radius is more extended for slower
rotating bars and for more centrally concentrated mass distributions
(see Fig. 6 and 7 in the above mentioned paper). Due to its high gaseous 
density, this central region will thus be the locus of high star
formation, and therefore the mean stellar age in there will be
considerably lower than that of the region immediately surrounding it.
The observations also
show the same central minimum in age and thus the simulations give
a good {\it qualitative} representation of this feature and allow us
to explain its formation and properties.

We have examined the mean stellar age profile for all the 
galaxies in our observational sample, and find that a central drop is
similarly 
seen in many (but not all) of the barred galaxies, while  
unbarred galaxies basically show little or no  
such central features. We similarly examined about half a dozen simulations
and found similar results and also could witness the formation of this
dip with time. Extensive comparisons with simulations covering a
broader spectrum of different galactic types would be able to 
provide better constraints on models of bar formation and a better
understanding of bar-driven secular evolution of galaxies.

Beyond this innermost region, up to one normalised length unit
(i.e. roughly up to the end of the bar) the mean age decreases with
increasing 
distance from the centre for both simulated and observed radial profiles. In the simulation, the disc forms inside-out, 
in good agreement with many observations, and this leads to a
negative age gradient. Furthermore, any gas in this region will be
pushed inwards by the bar to the region within the ILR and therefore
only few new stars will be born, so that any stellar profile
at formation will not be modified much at later times. 

We also note that the slope
of this radial density profile beyond the central dip is not 
constant with distance from the centre. In the region between
the maximum age to about half the bar length, the profile
is rather flat. For larger radii, however, the decrease is much
stronger, until a minimum age is reached. This behaviour is very clear
in the profile from the CALIFA data, both for 9 and
13 Gyr cases. It is also present for the simulation age profile,
although less dramatic.
The extent of the
two parts, i.e the flatter and the steeper decreasing parts, is also
in rather good agreement between observations and simulations.

At the end of this radial range where the age is decreasing, we have a
minimum. Again this is much clearer in the profiles obtained from
observations, but it can also be seen well in the 13 Gyr simulation
profile and, more as a change of slope, for the 9 Gyr one. This
difference between observations and simulations is due to a difference
in the corresponding morphologies (Fig.~\ref{fig:simprofile}). Namely
in NGC 5000 the bar is surrounded by an inner ring, i.e. a structure
which is usually gas-rich and a locus of intense star formation \citep[for a review, see][]{buta1996}. We thus expect a concentration of
young stars and a minimum in the radial age profile. Such inner rings
can be seen in a number of other simulations run by one of us (EA) and
have the expected properties. However the simulation snapshots analysed
here have a clear inner ring only in the gas and young populations (not
shown here), but not in the older populations where it has a more
lens-like component. Thus there should be less star formation in the simulation and the
mean age at these radii should not be much decreased with respect to
their surroundings.  


\section{Discussion}
\label{sec:discussion}

\subsection{How often do we see a central star formation enhancement?}

Our sample consists of 58 nearby galaxies, including 41 
isolated galaxies with regular morphologies (17 
barred and 24 unbarred), and 17
mergers or paired galaxies. A third of all the galaxies (19/58)
in our sample present a significant central upturn 
in $\log$\ewhae, including 13
barred isolated galaxies and 6 mergers/pairs. 
This is a large fraction compared to expectations 
from the traditional view that
spiral galaxies host redder and older 
galactic centers with less star formation than their outer discs. 
This result echoes a similar fraction of 
``turnover'' galaxies found in \citet{lin2017a}, where 17 out 
of 57 galaxies were identified as having a central turnover in 
\dindex. We note, however, that \citet{lin2017a} excluded 
mergers/pairs from their sample. If mergers and pairs are also 
excluded from our sample, the fraction of upturn 
galaxies remains similar though slightly higher, 
31.7\% (13/41). 

As discussed in \citet{lin2017a}, almost all of the 
galaxies with a central drop in
\dindex\ present central upturn features in both \ewhda\ and
$\log$\ewhae. Therefore, a higher upturn fraction in our sample 
should not be attributed to the fact that \citet{lin2017a}
used \dindex\ as the nominal definition of enhanced central 
star formation rather than $\log$\ewhae.
As pointed out above, by selection the EDGE-CALIFA
sample is biased to gas-rich galaxies. Therefore, a higher 
fraction of upturn galaxies in our sample might imply 
that the central star formation enhancement happens more
frequently in galaxies with more cold gas. As pointed out
earlier (\S\ref{sec:sample}), our 
sample is limited to relatively high-mass galaxies with few 
below $M_\ast\sim 10^{10}M_\odot$ (see Fig.~\ref{fig:sample1}).
Therefore, one might expect the upturn galaxy fraction to 
increase if the sample includes galaxies of lower masses which 
are more gas-rich.

It is thus natural to conclude that the central star formation
enhancement occurs commonly in local galaxies, with a fraction
of at least 1/3. The central upturn or turnover in the
spectral indices can be observed only when spatially resolved
spectroscopy like IFU data is available. 
 Multi-band imaging
should be also useful, and probably more applicable, if one
were to study such central features for larger samples.
In this case color indices can be used as reasonably good
indicators of recent star formation history, although they 
suffer from dust extinction more seriously than spectral indices.
We predict that a large fraction of galaxies must present 
a central drop in their color maps or profiles, an effect that 
can be tested with existing/future large-area imaging surveys 
at different redshifts.

\subsection{Can bars and mergers together fully account for the central star formation enhancement?}

The fact that all galaxies in our sample with a central
upturn in $\log$\ewhae\ are either barred, mergers, or pairs, 
and that no unbarred galaxies (24) present an upturn feature strongly suggests that bars, 
mergers and pairs fully account for the central star
formation enhancement occurring in our sample.
On the other hand, 
4 of the 17 barred galaxies and 11 of the 17
mergers/pairs present no upturn features in their centers.
This result suggests that the presence of a bar or tidal 
interactions is necessary, but not a sufficient condition for
the central star formation enhancement, in agreement with the conclusions of \citet{lin2017a}.

Our work  extends their findings 
by including mergers and paired galaxies in the analysis. 
Bars and mergers appear to respectively account for $\sim$2/3 and
$\sim$1/3 of the central upturn phenomenon. 
 Previous studies
of large samples from SDSS have examined the correlation of 
central star formation with both galaxy-galaxy interactions 
\citep[e.g.][]{li2008} and the internal bar structure 
\citep[e.g.][]{wang2012}. \citet{li2008} found $\sim40$\% of
the most strongly star-forming galaxies in the local Universe
have a close companion, while \citet{wang2012} found that only 
half of the galaxies with centrally enhanced star formation
 host a bar. The two studies combine to  suggest that 
bars and interactions can roughly account for the central star
formation enhancement occurring in low-redshift galaxies. 
It is encouraging that the same conclusion is reached by 
both the SDSS-based studies of large samples, i.e. 
\citet{li2008} and \citet{wang2012} which used 
single-fiber spectroscopy, the work of \citet{lin2017a}, and the current work which uses
integral field spectroscopy.

In addition to instabilities driven by bars and tidal interactions,
other mechanisms may be invoked, such as driving by spirals or asymmetries,
although presumably these may be much less efficient.
A common purpose of all these 
mechanisms is to transport cold gas from the disk to the central
kiloparsec, where star formation is triggered due to increased
gas density.  In this work we have shown that molecular gas
is indeed more concentrated when central star formation enhancement
is observed (Fig.~\ref{fig:strengthvsco}). More importantly, our work shows that the star
formation enhancement can be substantially explained by bars 
and mergers, with no need to have additional mechanisms. 
Again, however,
we should emphasize that our 
sample is biased to relatively high-mass and gas-rich 
galaxies. Larger samples covering wider ranges of mass and color
would be needed if one were to have a complete picture of the
physical mechanisms behind the central star formation enhancement.

\subsection{Bar-driven central SF enhancement as a long-lived effect}

Our results suggest that bar-induced central star 
formation is a long-term process lasting at least 1-2 Gyr.
This can be seen from 
both Fig.~\ref{fig:tstrength_hists} and 
Fig.~\ref{fig:strengthvsco}, where barred galaxies with
a central upturn in $\log$\ewhae\ also show a central upturn in
\ewhda\ and a central drop in \dindex.
We know that a central drop in \dindex\ indicates that
a considerable fraction of young stellar populations 
were formed in the central region 1-2 Gyr ago, 
a central upturn in \ewhda\
reveals the existence of a starburst ending 0.1-1 Gyr ago,
and a central upturn in $\log$\ewhae\ 
indicates ongoing star formation. Therefore, the fact that 
both a \dindex\ drop and an \ewhda\ upturn are associated
with barred, $\log$\ewhae-upturn galaxies implies that the central 
star formation induced by the bar  started at least 1-2 
Gyr ago, and is still happening at the moment.

\citet{krumholz2015} showed using simulations that the timescale between
bar-induced gas accumulation and 
a subsequent central starburst is only 10-20 Myr, after which the central region needs to 
accumulate more gas. 
The three SFH indicators used in the current work cannot tell whether the SF process is 
continuous or bursty, if the cycle timescale is only 10-20 Myr. 
What we can say is that there are young stellar populations; 
some are younger than a few Myr, and some are younger than 1-2 Gyr.
We find that the upturn and turnover strengths are correlated for barred galaxies,
with correlation coefficients ranging from 0.50 to 0.54, which supports the conclusion that 
central SF enhancement is long-lived. The correlations are weaker for merger/pair galaxies, and 
negligible for unbarred galaxies.

\subsection{Is high $c_\mathrm{mol}$ a necessary and sufficient condition for central star formation enhancement?} 

Our results show that high gas concentration is 
neither a necessary, nor sufficient condition for enhanced
central star formation to occur. This can be seen from both
Fig.~\ref{fig:tstrength_hists} and Fig.~\ref{fig:strengthvsco}.
On the one hand, the upturn galaxies (particularly those with a bar)
span the full range of $\log c_\mathrm{mol}$, although the
majority of them have higher gas concentration than non-upturn
galaxies, with $\log c_\mathrm{mol}\gtrsim 0.3$ in most cases. 
This result suggests that a highly concentrated gas distribution
is an important, but not a necessary condition for the central 
upturn to occur in our galaxies. 
On the other hand, we see some galaxies in our sample with high gas concentration $\log c_\mathrm{mol}>0.3$ but 
without a central upturn. This result indicates that high gas
concentration alone is not sufficient for the central upturn.

It is interesting that all of the galaxies with no central upturn
but with high gas concentration are in the category of mergers/pairs,
except for one galaxy, NGC7819, which is an unbarred galaxy
with $\log c_\mathrm{mol}\sim 0.6$ (see \S~\ref{sec:individuals}
for more discussion on this galaxy). For barred galaxies, 
we see that all those with $\log c_\mathrm{mol}>0.3$ present
a central upturn in $\log$\ewhae, with no exception, although a central
upturn/turnover in \ewhda\ or \dindex\ is not associated in a few
cases. Thus, bars seem to be more efficient than mergers in triggering
enhanced star formation in galactic centers, even when cold
gas is not highly concentrated. 
However, star formation may be underestimated in mergers due to high levels of dust extinction.
We conclude that high gas concentration is neither necessary, 
nor sufficient, and that the presence of a bar in most cases 
or mergers/interactions in other cases appear to be a crucial 
condition for central star formation enhancement.

\section{Conclusions and future work}
\label{sec:conclusions}

We have studied the spatially-resolved molecular gas and indicators of recent star-formation history
for 58 nearby galaxies using CARMA EDGE CO $J=1\rightarrow 0$ and 
CALIFA optical IFU data.
We divide our sample of 58 galaxies into three subsamples based on morphology: barred (17), unbarred (24), and mergers/pairs (17). 
The resolved gas and star formation history data are used to compare these three subsamples.
We use the equivalent width of the H$\alpha$ emission line $\log$\ewhae, 
equivalent width of the H$\delta$ absorption line \ewhda\ and the 4000 \AA~
break \dindex~to measure the strength of recent star formation in the central 
region compared to the outer part of the central region (inside the spiral arms). 
These three parameters allow us to probe the star formation history 
at three times: 0-30 Myr (\ewhae), 0.1-1 Gyr (\ewhda), and 1-2 Gyr (\dindex).
We measure a molecular gas concentration index $c_\mathrm{mol}$ 
defined as the ratio of the optical half-light radius to the molecular gas half-mass radius $r_\mathrm{50,mol}$ measured from radial profiles of the publicly available EDGE CO $J=1\rightarrow 0$ maps.

After comparing the central star formation history and molecular 
gas concentration for subsamples of barred, unbarred and merging/paired
galaxies, we reach the following conclusions: 
\begin{itemize}
  \item Out of the 58 galaxies in our primary sample, 19
  show a central upturn in $\log$\ewhae, of which 13 are barred, none are 
  unbarred, and 6 are mergers/pairs. Galaxies with upturns 
  have higher gas concentrations than barred or unbarred galaxies without upturns (Table~\ref{tab:table1}). 
  Merger/pair galaxies without upturns have average concentrations similar to galaxies with upturns.
  
  \item The level of enhanced central star formation is positively correlated with molecular gas concentration for barred galaxies, and in two out of the three SFH indicators. No significant correlations are found for unbarred or merger/pair galaxies (Figure~\ref{fig:strengthvsco}).

  \item Barred galaxies with upturns in $\log$\ewhae\ have significantly higher values of 
  upturn and turnover strengths than merger/pair galaxies with upturns (Table~\ref{tab:table1}). The average gas concentrations are 
  consistent between these two categories. However, barred galaxies with no 
  upturn have significantly lower gas concentrations than merger/pair galaxies 
  with no upturn.
  These results imply that bars are efficient in enhancing central
  star formation, which is long-lived (at least 1-2 Gyr).
  
  \item Our observational results provide strong support to the current
  theory of bar formation in which bars form and grow from inside out. Furthermore, they transport cold gas from the disk to the central region, which leads to 
  significant enhancement in star formation. 
  We compared our results with those from a simulation snapshot (\S\ref{sec:simresults}) and
found that the latter  
  successfully
  reproduce two major features in the azimuthally-averaged radial profiles 
  of mass-weighted age obtained from the data, namely the sharp 
  decrease of stellar age in the galactic center and the gradual decrease 
  of age with increasing distance from center. This qualitative comparison
  provides evidence for a picture in which cold gas is 
  transported inward due to a bar or tidal driving, which leads to the 
  growth and rejuvenation of the central region.
  
\end{itemize}

Some important
issues are not yet addressed in the current work, such as the 
correlation of central upturn strength with bar length and ellipticity (which are commonly used to quantify the strength of a bar), and
the distribution of cold gas and star formation indicators along 
and surrounding the bar. \citet{lin2017a} found a weak 
correlation between the radius of \dindex\ turnover and bar length.
It would be interesting to examine whether the gas concentration
is also somehow correlated with bar properties. Secondly, the merger/pair
category may be studied in more detail, e.g. by further splitting the 
galaxies into subsets according to pair separation and merger 
status to examine how the central star formation and
gas distribution evolve as the interaction/merger proceeds. 
It would also be interesting to have more detailed 
analyses of the exceptional galaxies in our sample, as mentioned
in the previous subsection. What are the reasons for variations 
in the barred galaxies? Do different concentrations indicate 
different stages of bar-driven gas transport? Finally, one may 
also want to examine the effects of different environments, such
as ram-pressure stripping and tidal stripping that occur in/around
massive dark halos and can effectively strip hot/cold gas of 
satellite galaxies. For this purpose our sample is probably too
small, and larger samples with both integral field spectroscopy 
and CO intensity mapping are needed.  

From the theoretical side, it is important to make comparisons
for different types of galaxies and for models with different
bar strengths, gas properties, feedback, star formation, as and dark matter
halo properties. In this work we have focused our observational 
analysis on stellar populations, while ignoring dynamical 
properties of our galaxies which can be measured from integral
field spectroscopy as well. As mentioned above, the radius where the central
upturn/turnover occurs may be related to the mass distribution
in the innermost region and the pattern speed of the bar
\citep[e.g.][]{athanassoula1992b}. The latter can be measured
by dynamical modelling of the kinematics of stars in the galaxy
based on integral field spectroscopy data. Therefore stellar population
synthesis and dynamical modelling in combination are expected to
provide more powerful constraints on bar formation models.

\bsp
\label{lastpage}

\section*{Acknowledgements}
This work is supported by the
National Key R\&D Program of China
(grant Nos. 2018YFA0404502), the National Key Basic Research Program of China (No. 2015CB857004), and the National Science Foundation of China (No. 11233005, 11325314, 11320101002, 11733004). RC acknowledges the support of McMaster University, a Mitacs Globalink Research Award (IT10717), and the China Scholarship Council. EA thanks the CNES for financial support. This work was granted access to the HPC resources of CINES under the allocations 2017-A0020407665 and 2018-A0040407665 attributed by GENCI (Grand Equipement National de Calcul Intensif), as well as the HPC resources of Aix-Marseille University financed by the project Equip@Meso (ANR-10-EQPX-29-01) of the program
``Investissements d'Avenir'' supervised by the Agence Nationale de la Recherche. HJM acknowledges the support from NSF AST-1517528.
The research of CDW is supported by grants from the Natural Sciences and Engineering Research Council of Canada and the Canada Research Chairs program. LL was supported by the National Science Foundation of China (No. 11703063).

This study uses data provided by the CARMA Extragalactic Database for Galaxy Evolution (EDGE) survey (\url{http://www.astro.umd.edu/EDGE/}), the Calar Alto Legacy Integral Field Area (CALIFA) survey (\url{http://califa.caha.es/}), the NASA-Sloan Atlas (\url{http://www.nsatlas.org/}), the Sloan Digital Sky Survey (\url{https://www.sdss.org/}), the HyperLeda database (\url{http://leda.univ-lyon1.fr}), and the SIMBAD database, operated at CDS, Strasbourg, France.
CALIFA is based on observations collected at the Centro Astron\'omico Hispano Alem\'an (CAHA) at Calar Alto, operated jointly by the Max-Planck-Institut f\"ur Astronomie and the Instituto de Astrof\'isica de Andaluc\'ia (CSIC).

\bibliographystyle{mnras}
\bibliography{main.bib}

\appendix

\section{Tables of galaxy properties}

Table~\ref{tab:tableA1} shows basic properties of the galaxies in our sample,
and Table~\ref{tab:tableA2} shows measurements derived from our analysis.

\begin{table*}
	\centering
	\caption{Basic properties of the galaxies in our sample. The full version is available in machine-readable format. }
	\label{tab:tableA1}
	\begin{tabular}{lrrrrrrrrrr} 
\hline 
 Galaxy & No. & R.A.    & Dec.    & $z$ & $\log M_*$   & NUV-$r$ & $r_{50}$ &  $\log M_\mathrm{HI}$ & $\log M_\mathrm{H_2}$ & Type \\ 
        &     & (J2000) & (J2000) &     & $(M_\odot)$  & (mag)   & (\arcsec) &  $(M_\odot)$          & $(M_\odot)$           &      \\ 
 (1) & (2) & (3) & (4) & (5) & (6) & (7) & (8) & (9) & (10) & (11)  \\
\hline 
 & & & & & Barred & & & & &  \\ 
\hline 
IC1683 & 2 & 20.66220 & 34.43713 & 0.016 & 10.4 & 4.1 & 13.1 & 9.2 & 9.5 & Sb \\ 
\hline 
 & & & & & Unbarred & & & & &  \\ 
\hline 
NGC7819 & 21 & 1.10211 & 31.47201 & 0.017 & 10.1 & 2.6 & 23.8 & 9.6 & 9.2 & Sc \\ 
\hline 
 & & & & & Merger/pair & & & & & \\ 
\hline 
NGC0523 & 45 & 21.33649 & 34.02495 & 0.016 & 10.6 & 4.1 & 24.2 & 10.1 & 9.5 & Sd \\ 
\hline
\multicolumn{11}{l}{(1): Galaxy name.}\\
\multicolumn{11}{l}{(3): Right ascension (degrees), from SDSS DR7.}\\
\multicolumn{11}{l}{(4): Declination (degrees), from SDSS DR7.}\\
\multicolumn{11}{l}{(5): Raw redshift measured from the CALIFA datacubes.}\\
\multicolumn{11}{l}{(6): Stellar mass from CALIFA DR3 reanalysis of SDSS DR7 \textit{ugriz} growth curves \citep{walcher2014}.}\\
\multicolumn{11}{l}{(7): \nuvr~magnitude from the NASA-Sloan Atlas (NSA).}\\
\multicolumn{11}{l}{(8): $r$-band half-light radius $r_{50}$ from the NSA.}\\
\multicolumn{11}{l}{(9): Neutral hydrogen mass from ALFALFA 100\% catalog \citep{haynes2018} where available, or from the HyperLeda database.}\\
\multicolumn{11}{l}{(10): The total detected H$_2$ mass in the whole CO image.}\\
\multicolumn{11}{l}{(11): The morphological type (RC3) provided in CALIFA DR3.}\\
\end{tabular}
\end{table*}

\begin{table*}
	\centering
	\caption{Quantities derived from spatially-resolved optical IFU and molecular gas maps. The full table is available in machine-readable format.}
	\label{tab:tableA2}
	\begin{tabular}{lrrrrrrrrrr} 
\hline 
Galaxy & Upturn? & $\Delta$D$_\mathrm{n}$(4000) & $\Delta$EW(H$\delta_A$) & $\Delta \log$EW(H$\alpha$)  &  $r_\mathrm{50,mol}$ & $\log c_\mathrm{mol}$ & $\log \tau_\mathrm{depl.}$ & $\log \tau_\mathrm{depl.}$ & $\log \tau_\mathrm{cen.}/\tau_\mathrm{disk.}$  \\ 
 & & & & & & & Center & Disk & \\ 
 & & & (\AA) & & (\arcsec) & & (yr) & (yr) & \\ 
 (1) & (2) & (3) & (4) & (5) & (6) & (7) & (8) & (9) & (10)   \\
\hline 
\multicolumn{10}{c}{Barred} \\
\hline 
IC1683 & Y & $0.66$ & $1.69$ & $0.58$ & $5.63$ &  $0.37$ & $8.95\pm0.07$ & $9.25\pm0.06$ & $-0.30\pm0.09$ \\ 
\hline 
\multicolumn{10}{c}{Unbarred} \\
\hline 
NGC7819 & N & $0.02$ & $0.70$ & $0.31$ & $5.96$ &  $0.60$ & $8.85\pm0.07$ & $8.92\pm0.06$ & $-0.06\pm0.09$ \\ 
\hline 
\multicolumn{10}{c}{Merger/pair} \\
\hline
NGC0523 & N & $-0.13$ & $-0.07$ & $0.10$ & $18.05$ &  $0.13$ & $9.87\pm0.10$ & $9.29\pm0.06$ & $0.57\pm0.12$ \\ 
		\hline
\multicolumn{10}{l}{Note: galaxies that are not in the reduced sample (\S\ref{sec:sample}) have values of ``n/a'' in columns 8-10.}\\
\multicolumn{10}{l}{(1): Galaxy name.}\\
\multicolumn{10}{l}{(2): Does this galaxy have an upturn in $\log$\ewhae (\S\ref{sec:turnoverid})? (Y or N).}\\
\multicolumn{10}{l}{(3): \dindex\ turnover strength.}\\
\multicolumn{10}{l}{(4): \ewhda\ upturn strength.}\\
\multicolumn{10}{l}{(5): $\log$\ewhae\ upturn strength.}\\
\multicolumn{10}{l}{(6): Molecular gas half-mass radius.}\\
\multicolumn{10}{l}{(7): Molecular gas concentration index (Eq.~\ref{eq:cmol}).}\\
\multicolumn{10}{l}{(8): Molecular gas depletion time (Eq.~\ref{eq:tdepl}) in the ``center'' (0.5 kpc semi-major axis).}\\
\multicolumn{10}{l}{(9): Molecular gas depletion time in the ``disk'' (between 0.5 kpc and $r_{50}$).}\\
\multicolumn{10}{l}{(10): Center-to-disk depletion time ratio.}\\
\end{tabular}
\end{table*}

\subsection{Notes on individual galaxies}
\label{sec:individuals}

\begin{figure*}
\centering
\includegraphics[width=0.9\textwidth]{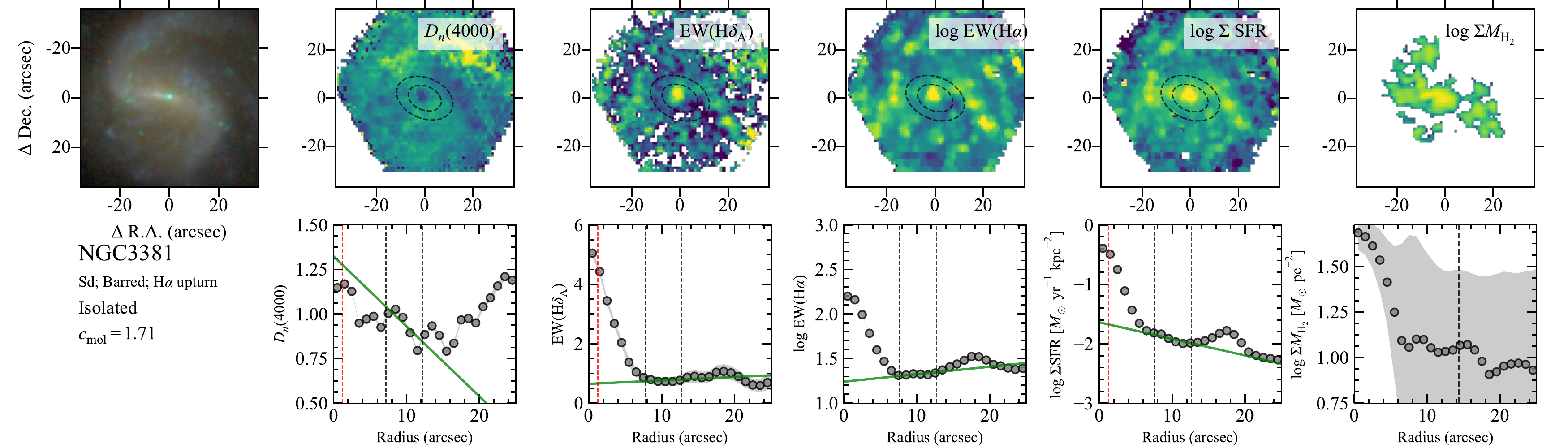}
\includegraphics[width=0.9\textwidth]{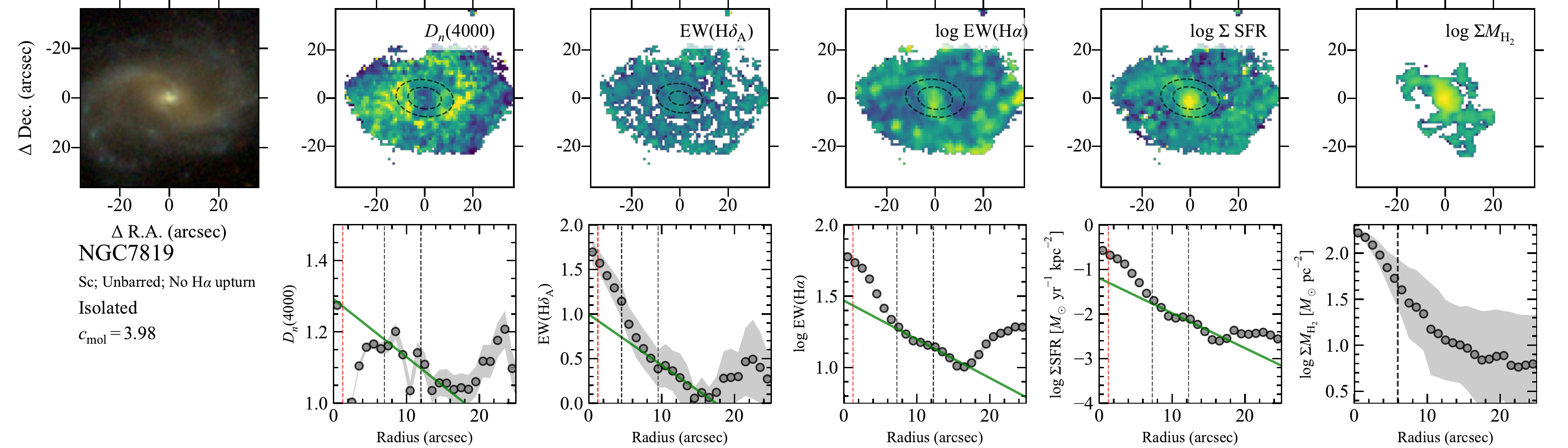}
\includegraphics[width=0.9\textwidth]{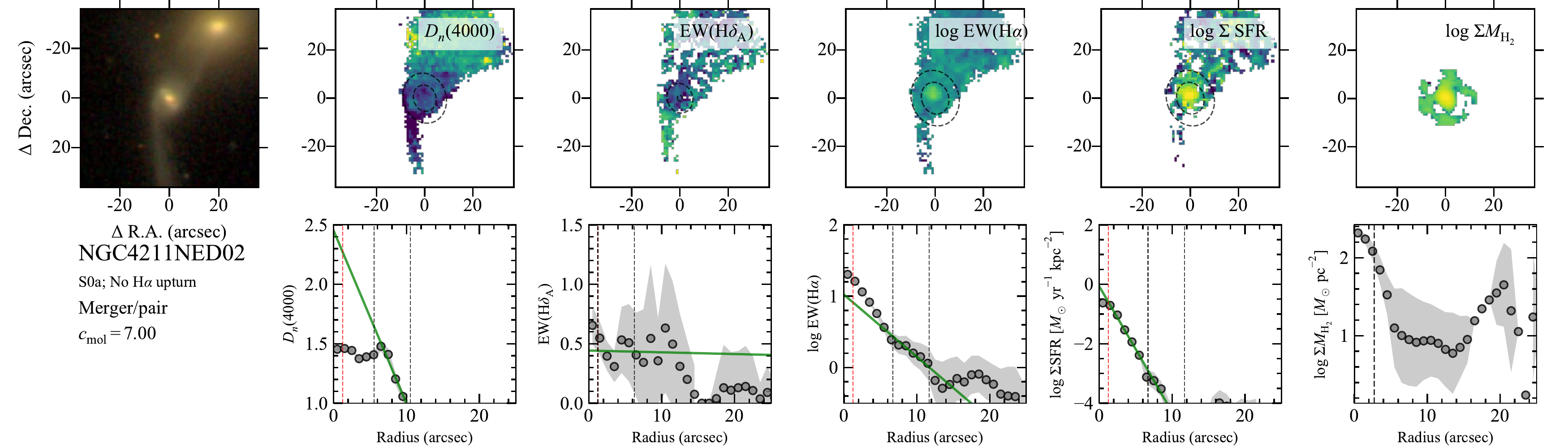}
\caption{Maps and radial profiles of additional example galaxies in our sample. See caption of Fig.~\ref{fig:mapsandprofiles1} for description of quantities.}
\label{fig:mapsandprofiles2}
\end{figure*}

Here we mention a few galaxies which show unusual or extreme behaviours
based on Figure~\ref{fig:strengthvsco}. Their optical images, SFH indicators, molecular gas maps and radial profiles are shown in Fig.~\ref{fig:mapsandprofiles2}.

\begin{description}

    \item[NGC3381:] the barred galaxy with a very high $\Delta $\ewhda\ 
    but with only intermediate $c_\mathrm{mol}$. This galaxy is 
    blue in colour (\nuvr\ $=2.0$, the bluest of all our barred galaxies),
    and has a low stellar mass ($10^{9.6}M_\odot$; the lowest of all our
    barred galaxies). This galaxy has an observed value of \ewhda$=5.05$\AA,
    the highest in our sample (middle panel of
    Fig.~\ref{fig:obs_extrap_scatter}), and is consistent with the
    definition of a post-starburst galaxy.

    \item[NGC7819:] the unbarred galaxy with the highest $c_\mathrm{mol}$.
    With an \nuvr\ of 2.1 mag (the bluest of the unbarred galaxies), this
    galaxy is clearly blue and star-forming. This galaxy has quite a large
    optical radius (23.8\arcsec), and a relatively small  $r_\mathrm{50,mol}$. Given this high concentration, why does it not 
    show enhanced central star formation in any of our SFH indicators? 
    With a central $\log \Sigma_\mathrm{H_2}$ of $2.27 M_\odot$pc$^{-2}$ 
    and central $\log \Sigma_\mathrm{SFR}$ of $-0.58 M_\odot$yr$^{-1}$kpc$^{-2}$,
    which are not dissimilar from the central values of these quantities in
    barred upturn galaxies (Fig.~\ref{fig:ks}), it appears that central SFR
    and H$_2$ surface densities being similar to barred galaxies, and having
    a high molecular gas concentration are not sufficient conditions for
    enhanced central star formation. A bar or galaxy interaction appears to
    be needed too, as discussed above. It would be interesting to study the
    dense gas in this system -- it could be that the gas is not sufficiently
    dense to form stars without a bar.
    
    \item[NGC4211NED02:] the merger galaxy with the highest $c_\mathrm{mol}$
    but no significant central upturn. 
    This galaxy has \nuvr\ of 3.7, and stellar mass of $10^{10.0}M_\odot$.
    It is in a close pair with NGC4211A (not in our sample), with a
    projected separation of approximately 16 kpc. The CO emission for this
    galaxy is very compact, and since $r_\mathrm{50,mol}$ is less than twice
    the CARMA beam scale, $c_\mathrm{mol}$ should be considered a lower
    limit 
    (as indicated in Table~\ref{tab:tableA2}). This galaxy does not show an upturn, however it shows a significant \dindex\ turnover (see the bottom-right panel of 
    Fig.~\ref{fig:strengthvsco}). The turnover strength for this particular
    galaxy should be interpreted with caution, because the size of the
    central region of this galaxy is quite close to the resolution of
    CALIFA. Although the optical radius of this galaxy is 19\arcsec, this is
    likely due to the irregular morphology. The size of the inner region is
    much smaller, which makes our fitting less reliable.
    
\end{description}


\end{document}